\documentclass{emulateapj}
\usepackage{natbib}
\usepackage{graphicx}
\usepackage{amssymb}
\usepackage{epstopdf}
\usepackage{subfigure}

\usepackage{graphicx}
\usepackage{soul}

\usepackage{color}

\begin{document}

\title{Cosmic Ray Small Scale Anisotropies and Local Turbulent Magnetic Fields}

\author{V. L\'opez-Barquero$^{1}$, R. Farber$^{2}$, S. Xu$^{3}$, P. Desiati$^{4,5}$, A. Lazarian$^5$}

\affiliation{1. Department of Physics, University of Wisconsin, Madison, Wisconsin 53706, USA \\
2. Department of Physics and Astronomy, Wheaton College, Norton, MA 02766, USA\\
3. Department of Astronomy, School of Physics, Peking University, Beijing
100871, China\\ 
               4. Wisconsin IceCube Particle Astrophysics Center (WIPAC), University of Wisconsin, Madison, WI 53703, USA\\
               5. Department of Astronomy, University of Wisconsin, Madison, WI 53706, USA}


\begin{abstract}
Cosmic ray anisotropy has been observed in a wide energy range and at different angular scales by a variety of experiments over the past decade.
However, no comprehensive or satisfactory explanation has been put forth to date. 
The arrival distribution of cosmic rays at Earth is the convolution of the distribution of their sources and of the effects of geometry and properties of the magnetic field through which particles propagate. It is generally believed that the anisotropy topology at the largest angular scale is adiabatically shaped by diffusion in the structured interstellar magnetic field. On the contrary, the medium- and small-scale angular structure could be an effect of nondiffusive propagation of cosmic rays in perturbed magnetic fields. In particular, a possible explanation of the observed small-scale anisotropy observed at TeV energy scale, may come from the effect of particle scattering in turbulent magnetized plasmas. We perform numerical integration of test particle trajectories in low-$\beta$ compressible magnetohydrodynamic turbulence to study how the cosmic rays arrival direction distribution is perturbed when they stream along the local turbulent magnetic field.  We utilize Liouville's theorem for obtaining the anisotropy at Earth and provide the theoretical framework for the application of the theorem in the specific case of cosmic ray arrival distribution. In this work, we discuss the effects on the anisotropy arising from propagation in this inhomogeneous and turbulent interstellar magnetic field.

\end{abstract}


\section{Introduction} 
\label{sec:intro}

Cosmic rays are found to possess a small but measurable anisotropy in their arrival direction distribution at Earth. The origin of the observed anisotropy is not yet understood. However, it is reasonable to assume that it is a combination of effects correlated to the distribution of the galactic sources of cosmic rays, the geometry and turbulence properties of the galactic magnetic field, and the propagation in interstellar magnetized plasmas. These are likely to be responsible for the complex shape of the energy spectrum as well~\citep{gaisser_2013}. Since we don't know the locations of cosmic ray sources and we lack details of the interstellar magnetic field, understanding these observations of anisotropy is not an easy task.

Above the energy range where cosmic rays are directly affected by inner heliospheric processes (see, e.g.,~\cite{florinski_2013,manuel_2014,florinski_2015}), a statistically significant anisotropy has been observed by a variety of experiments, sensitive to different energy ranges (from tens of GeV to a few PeV), located on or below the Earth's surface in the Northern Hemisphere~\citep{nagashima_1998,hall_1999,amenomori_2005,amenomori_2006,guillian_2007,abdo_2009,aglietta_2009,zhang_2009,munakata_2010,amenomori_2011,dejong_2011,shuwang_2011,bartoli_2015} and in the Southern Hemisphere~\citep{abbasi_2010,abbasi_2011,abbasi_2012,aartsen_2013}.

The global anisotropy changes with energy in a non-trivial fashion. From about 100 GeV to tens of TeV, it has been observed to have an approximately consistent structure at the largest scale, although its amplitude appears to increase with energy. Above a few tens of TeV, the observed progressive change in the anisotropy topology may indicate a transition between two processes shaping the particles' arrival distribution at Earth. The observation could be qualitatively explained on the basis of diffusive propagation of cosmic rays in the Milky Way from stochastically distributed sources, responsible for generating a gradient in cosmic ray density. 

Numerical studies of particle propagation in a scenario of homogeneous and isotropic diffusion in the Galaxy predicts that the cosmic ray density gradient, and the consequent induced anisotropy, has a dipole shape with direction toward the source of the particle and with amplitude that directly depends on the diffusion coefficient. In particular, for a given realization of galactic source spatial distribution, the dipole anisotropy would point toward the source with the largest contribution~\citep{erlykin_2006,blasi_2012,ptuskin_2012,pohl_2013,sveshnikova_2013,savchenko_2015}, which may change with energy, in agreement with observations. On the other hand, the systematic overestimation of the anisotropy amplitude may be partially compensated by the fact that diffusion is expected to be anisotropic (see, e.g.,~\cite{effenberger_2012}), thus modifying the expected cosmic ray density gradient shape as a function of the source direction with respect to the regular galactic magnetic field~\citep{kumar_2014}. The misalignment between the cosmic ray density gradient and the regular galactic magnetic field would prevent pointing to any specific source, and it would suppress the anisotropy amplitude to a value closer to what has been observed~\citep{mertsch_2015}.

Another scenario is that it is the transition from heliospheric- to interstellar-dominated contributions, starting at a 10 TeV energy scale and culminating around 100-200 TeV, at the origin of the shift in anisotropy. In~\cite{desiati_lazarian_2013}, this scenario was proposed noting that 10 TeV protons have an average gyroradius, in a $\mu$G scale magnetic field, on the order of the thickness of the heliosphere (see, e.g.,~\cite{pogorelov_2006}). In addition, the dynamical instabilities of the heliospheric magnetized plasma at smaller scales may generate strong scattering that redistributes the arrival direction of TeV cosmic rays. 
This heliospheric scenario is studied and presented in our companion paper~\cite{barquero_2016}.
Such strong scattering may be able to produce large localized particle gradients, experimentally interpreted as medium- or small-scale anisotropy. In~\cite{schwadron_2014}, a scenario of weak influence of the heliospheric magnetic field on TeV protons was explored, thus interpreting the observations as directly related to the ordering of the local interstellar magnetic field (see also~\cite{zhang_2014}).

%
The anisotropy appears to possess a complex angular structure with evidence of a harder cosmic ray spectrum within the localized excess region in the apparent direction of the heliospheric tail~\citep{amenomori_2007,abdo_2008,bartoli_2013,abeysekara_2014}. The decomposition of TeV cosmic ray anisotropy in the individual spherical harmonic contributions shows that the arrival direction distribution is dominated by large-scale structures (such as, e.g., dipole and quadrupole) with a relative intensity on the order of 10$^{-3}$, but that medium- and small-scale angular structures are significantly contributing with relative intensities below 10$^{-4}$.

The experimental determination of small angular scale anisotropy is generally performed by filtering out the large-scale modulations from the observed 
arrival direction distribution, thus retaining all structures with large angular gradients. Some of the small-scale anisotropy features seem to be correlated to regions in the sky where the global anisotropy has large variations (see~\cite{desiati_lazarian_2013}) or may be an effect of re-acceleration by magnetic reconnection processes in the tail of the heliosphere~\citep{lazarian_desiati_2010,desiati_lazarian_2012}. However, globally, the observed small scale anisotropy may appear to be rather randomly structured and, therefore, possibly a natural consequence of cosmic ray propagation in the local turbulent magnetic field in the presence of a global anisotropy~\citep{giacinti_sigl_2012}. The global anisotropy, at all angular scales, arises from the same physical processes, thus it is impossible to disentangle its origin. However, as a first approximation, it is generally assumed that the anisotropy at the largest scale is dominated by global physical processes (such as a density gradient from sources of cosmic rays or from convective effects originated by large-scale cumulative stellar winds, and from propagation through the regular galactic magnetic field), while the small angular scales are dominated by local processes. 

A complex angular power spectrum is asymptotically generated by progressive decomposition of the energy of an initial anisotropy distribution (for instance a dipole) into higher multipoles, by the effect of scattering off magnetic turbulence~\citep{ahlers_2014,ahlers_mertsch_2015}. The conservation of phase space density, as stated by Liouville's theorem, predicts, for an idealized situation of a homogenous large-scale anisotropy, the total sum of the multipoles' angular terms is conserved. This makes it possible to generate small-scale structures, as shown in~\cite{ahlers_2014}.
%


Turbulence in astrophysical plasmas have significant effects on particle propagation, in that the stochastic nature of magnetic field lines is transmitted to the particles' trajectories. If particles are tied to magnetic field lines, the maximum perpendicular diffusion rate is set by the rate of perpendicular field line wandering, scaled by particle velocity (field line random walk). This has been extensively discussed by, e.g.~\cite{jokipii_1966,jokipii_parker_1969,rr_1978,giacalone_jokipii_1999,minnie_2009,shalchi_2009} in the context of diffusion at distance scale larger than the turbulence injection scale. 
%
%
%
%
On scales smaller than the injection scale, particles are characterized by super-diffusion in the perpendicular direction of the mean magnetic field. This behavior is described by the so-called Richardson diffusion, where particle separation grows as (time)$^{3/2}$~\citep{ly_2014}. In~\cite{eyink_2011} it was shown that this is directly connected to the separation of turbulent magnetic field lines, which grows as (distance)$^{3/2}$~\citep{lv_1999}. 
In this case, the stochastic nature of the astrophysical magnetic fields (such as, e.g. the interstellar magnetic field, or, at larger scales, the intercluster magnetic field), inevitably produces chaotic particle trajectories, meaning that the geometry of particle trajectories is highly sensitive to the actual initial conditions. Diffusion by magnetic field line wandering is found to be a dominant contribution and stronger than the extreme case of Bohm diffusion, where scattering frequency is one per particle gyration.
The important aspect that determines the properties of a large ensemble of particles is its statistical nature, which in this case is influenced by the properties of the magnetic field and specifically by the induced scattering rate. While individual trajectories may have chaotic properties and are, therefore, practically/realistically unpredictable, the large ensemble of particles can still be deterministically described. For these reasons, the recorded spatial distribution of the ensemble will be statistically determined and a direct consequence of the properties of the turbulent magnetic field.
In this work, a study of particle propagation in compressible magnetohydrodynamics (MHD) turbulence is performed in the context of its effects on arrival direction distribution.

In addition, depending on the degree to which magnetic field lines diverge on scales less than the particle gyroradius, pitch angle scattering on small-scale magnetic perturbations affects particle distribution at the large spatial scale, thus increasing the diffusion coefficient depending on the large geometrical scale of the magnetic field turbulence (see, e.g.,~\cite{desiati_zweibel_2014} and references therein). The anisotropic nature of interstellar turbulence and the properties of turbulence itself can significantly complicate the description of cosmic ray transport (see, e.g.,~\cite{yan_lazarian_2008,yan_lazarian_2011,beresnyak_2011}).
%



This paper is organized as follows. In section~\ref{sec:prop}, we describe the turbulent magnetic field used and the test particle trajectory integration. In section~\ref{sec:lt}, we discuss the validity of applying Liouville's theorem in the context of this work by statistically assessing the level of conservation of magnetic moment. The results of the  study are presented in section~\ref{sec:res} and discussed in section~\ref{sec:disc}.  Concluding remarks follow in section~\ref{sec:concl}.

%
\section{Cosmic ray propagation in turbulent magnetic fields}
\label{sec:prop}

Cosmic rays, which are accelerated in their sources and ``injected" into the interstellar medium (ISM), are free to propagate through the interstellar magnetic field. Globally, the galactic magnetic field is characterized by a large-scale regular component and a small-scale random component (see, e.g.,~\cite{jansson_farrar_2012a,jansson_farrar_2012b}). The regular component can be described as a superposition of spiral and toroidal structures, possibly with a contribution perpendicular to the galactic plane, and the random component represents the stochastic perturbations of the regular field caused by the dynamics of the galaxy and its density distribution. An important property of the interstellar magnetic field is turbulence. A variety of observations show that the coherence scale of turbulent magnetic fields is on the order of 10 pc in spiral arm regions (with more frequent stellar formation activity) and on the order of 100 pc in the interarm regions~\citep{haverkorn_2008}. The injection scale of turbulence is determined by the scale at which the magnetic perturbation is generated (for instance, by stellar collapse or binary mergers).

Astrophysical plasmas are typically highly ionized and have high Reynolds numbers, thus the dynamics of the flow is dominated by nonlinear convective processes at the largest scale. In such conditions, turbulence develops and magnetized Alfv\'enic eddies dynamically cascade to smaller scales and progressively elongate along the magnetic field lines, as initially proposed by~\cite{sg_1994,gs_1995} (see also~\cite{lazarian_2007} for a review). This model of incompressible MHD turbulence predicts a Kolmogorov-type energy power spectrum $E(k_{\perp})\propto k_{\perp}^{-5/3}$ in terms of the wave-vector component perpendicular to the local direction of the magnetic field, while the parallel component of the wave vector is $k_{\parallel} = k_{\perp}^{2/3}$.
%
%
In those pioneering papers, the theory assumes the injection of energy at scale $L$ and the injection velocity equal to the Alfv\'en velocity in the fluid $V_A$, i.e., the Alfv\'en Mach number $M_A\equiv (V_L/V_A) = 1$ (i.e., trans-Alfv\'enic turbulence), where $V_L$ is the plasma velocity at injection scale $L$. The model was later generalized for both sub-Alfv\'enic (i.e., $M_A < 1$) and super-Alfv\'enic (i.e., $M_A > 1$) cases~\citep{lv_1999,lazarian_2006} (see also~\cite{lazarian_2012,burkhart_2014}).
Typically, the ISM is characterized by $M_A \lesssim$ 1~\citep{gaensler_2011}.
This means that magnetic field lines do not typically fluctuate too far from the mean direction. 

While the model by~\cite{gs_1995} describes incompressible MHD turbulence, modeling compressible turbulence turned out to be more complex. In isothermal plasmas, there are three types of MHD waves: Alfv\'en, slow, and fast waves. Alfv\'en modes are incompressible, while slow and fast modes are compressible. The compressible modes are conjectured to resemble incompressible behavior at high-$\beta$ (i.e., high gas to magnetic pressure ratio)~\citep{lg_2001}; however, ISM plasmas are typically characterized by low-$\beta$. \cite{cho_lazarian_2002} investigated the scaling properties of low-$\beta$ compressible sub-Alfv\'enic MHD turbulence and confirmed that slow compressible modes behave as the Alfv\'en incompressible modes but also that the fast modes are isotropic, since their velocity does not depend on magnetic field direction.

The spatial distribution of particles propagating in a turbulent field is affected by the magnetic perturbations within the particle autocorrelation length scale (or mean free path). To study the correlation between turbulence and cosmic ray distribution, trajectories of test particles were integrated in an MHD turbulent magnetic field. The methodology used is described in the next two sections.


%
\subsection{Turbulent magnetic field}
\label{ssec:turbmf}

Test particle trajectories are integrated in a compressible sub-Alfv\'enic isothermal MHD turbulence in low-$\beta$ based on numerical calculations developed by~\cite{cho_lazarian_2002}.
In the model, turbulence is driven solenoidally in Fourier space, setting the velocity and density fields initially to unity. The calculation is performed in a cube with side $L_{box} = 512$ grid-points, with inertial range from $L_{inj}=204.8$ grid-points (i.e., $0.4\times L_{box}$) down to a ``damping" scale of $L_{min}=5$ grid-points.

The average magnetic field is directed along the x-axis and turbulence is characterized by a gas-to-magnetic pressure value of $\beta \sim 0.2$ and by Alfv\'enic Mach number $M_A = 0.773$. Such a Mach number corresponds to the fluctuations being on the order of the mean magnetic field at the injection scale, which is in agreement with what we would have expected from the local ISM. The external mean magnetic field is the only controlled parameter in this MHD model.


\subsection{Cosmic ray propagation}
\label{ssec:crprop}

The analysis is performed by integrating proton trajectories in a static magnetic field, using the set of 6-dimensional ordinary differential equations
\begin{eqnarray}
\frac{d\vec{p}}{dt} & = & q \left( \vec{u}\times\vec{B} \right)\\
\label{eq:motion1}
\frac{d\vec{r}}{dt}  & = & \vec{u},
\label{eq:motion2}
\end{eqnarray}
describing the Lorentz force and the particle velocity $\vec{u}$, where 
$\vec{r}$ is the particle position vector and $\vec{p}$ the momemtum. For $\vec{B}$, we use one steady state realization of the magnetic field described in section~\ref{ssec:turbmf}.
The equations are integrated using the Bulirsch-Stoer integration method with adaptive time step. At each integration step, the magnetic field is interpolated using a 3D cubic spline, and integration is stopped when particles cross the border of the MHD simulation box.
The accuracy of the particle trajectory integration for this study was assessed and is discussed in Appendix~\ref{app:a}.
The choice of one specific realization of the magnetic field is justified by the fact that particle velocity is much larger than the plasma Alfv\'en velocity, thus induced electric fields can be neglected.

The magnetic field can be interpreted as a snapshot of the local interstellar magnetic field. Since the MHD simulation can be scaled to an arbitrary injection scale, in this study, two physical scenarios are investigated: an injection scale of $L_{inj}$ = 10 pc, which is approximately the magnitude typical of the Milky Way spiral arms, and $L_{inj}$ = 100 pc, which is characteristic of the interarm regions (see section~\ref{sec:prop}). Although the ISM surrounding the solar system seems to have interarm properties~\citep{frisch_2012,frisch_2014},  both cases are taken into consideration. Assuming a particle gyroradius of 5 grid-points (corresponding to the smallest spatial scale in the MHD simulation), the injection scale sets the particle energy scale. In order to avoid grossly underestimating the effects of magnetic perturbations smaller than the particle gyroradius, the latter is set to 20 grid-points as well.
\begin{table}[!t]
\caption{Physics parameters of simulation sets}
\centering
\begin{tabular}{ lllllll }
\hline
Set	& $E_p$		& $r_L$	& $L_{inj}$	& $L_{mfp}$	&Particles			& $\langle B_0 \rangle$	\\
\hline
1	& 750 TeV	& 0.24 pc	& 10 pc		& 5.0 pc		&$3 \times 10^5$	& 3 $\mu$G	\\
2	& 7.5 PeV		& 2.4 pc	& 100 pc		& 50 pc		&$10^5$			& 3 $\mu$G	\\
3	& 30 PeV		& 10 pc	& 100 pc		& 60 pc		&$10^5$			& 3 $\mu$G	\\
\hline
\end{tabular}
\label{tab:sets}
\end{table}
Table~\ref{tab:sets} shows the three trajectory integration data sets used in this study. They cover an energy range from 750 TeV to 30 PeV, partially overlapping the recent cosmic ray observations reported by the IceCube Observatory~\citep{abbasi_2012,aartsen_2013,aartsen_2015}.

In order to study the effect of interstellar magnetic turbulence on the arrival direction distribution of cosmic rays, a large number of particles would need to be injected with randomly uniform directions on a spheric surface centered at the Earth and with a radius larger than the mean free path. Unfortunately, such method would be highly inefficient because a large fraction of the injected particles would never reach Earth. The alternative approach is typically to use the so-called ``back-propagation method," where 
the trajectories of particles with opposite charge are integrated from Earth, with initial directions uniformly distributed, backward into outer space. Since energy losses are negligible for proton particles, their energy is conserved, and therefore their trajectories can be time-reversed, provided there are no collisional scattering processes and no resonant scattering. Under such general conditions, Liouville's theorem states that particle density in phase space is conserved along particle trajectories. In the presence of turbulence, magnetic fields can vary faster in space than the particle gyroradius, thus breaking adiabaticity and effectively inducing collisional processes. In section~\ref{sec:lt}, the applicability of Liouville's theorem in the present case is discussed.

For set 1 of Table~$\ref{tab:sets}$, $3 \times 10^5$ proton trajectories are integrated, while for sets 2 and 3, the number of particles is $10^5$. The numerical trajectory integration starts from a point at the center of a 3$\times$3 MHD simulation box, with uniform direction distribution, and stops when particles reach the edge of the integration volume. The integration box corresponds to a distance scale of 75 pc (for set 1 in table~\ref{tab:sets}) and of 750 pc (for sets 2 and 3 in Table~\ref{tab:sets}). The back-propagated trajectories calculated in this way provide the information on the effects of the magnetic field on an initially uniform particle distribution emanating from one point. In other words, they provide a map of regions that are magnetically connected to the origin at a given distance. The calculation implicitly takes into account the dynamical processes of a particle's motion in a magnetic field, including drifts and pitch angle scattering. Under the hypothesis that such trajectories are time-reversible (see section~\ref{sec:lt}), they can be interpreted as directly propagating from outer space  back to the original point (assumed to coincide with Earth). This means that the particle distribution far from Earth, resulting from the back-propagation numerical calculations, corresponds to a perfectly uniform arrival distribution at Earth.

The numerically calculated trajectories can be used, therefore, to determine the arrival distribution at Earth as a consequence of a global anisotropy at a large distance. Such a global anisotropy, which is the effect of a particle density gradient induced by sources of cosmic rays or by convective processes, is treated as a weight on the forward-inverted trajectories, and it effectively produces a nonuniform arrival direction distribution at Earth, which is described in section~\ref{sec:skymaps}.

\section{The validity of Liouville's theorem}
\label{sec:lt}

%
Generating a large number of particle trajectories that pass through a point in space is implicitly highly inefficient, since most particles will bypass the target point. One possibility is to increase the size of the target to record those particles that pass ``nearby," or to appeal to Liouville's theorem.
%

Liouville's theorem states that the density of particles in the neighborhood of a given system in phase space is constant if restrictions are imposed on the system~\citep{goldstein_2002}, as shown below.
To determine its validity for our specific case of cosmic ray arrival distribution, we shall start with the continuity equation in phase space~\citep{bradt_2008a,bradt_2008b}, under the assumption that the number of particles stays fixed:
\begin{equation}
\frac{\partial \rho }{\partial t} = -\vec\nabla \cdot ( \rho \vec{v}) -\vec\nabla_{p} \cdot ( \rho \vec{F})
\end{equation}
where $\vec\nabla_{p}$ is the {\tt del} operator in momentum space, $\rho$ the distribution function, and $\vec F$ the applied external force. 

Upon expansion of this expression, we arrive at: 
\begin{equation}
\label{continuity}
\frac{\partial \rho }{\partial t} = - \rho \vec\nabla \cdot (\vec{v}) -\vec{v}  \cdot \vec\nabla ( \rho )  - \rho \vec\nabla_{p} \cdot (\vec{F}) -\vec{F}  \cdot  \vec\nabla_{p}( \rho)
\end{equation}

The first term on the right-hand side of Eq. $\ref{continuity}$ vanishes, since the divergence of the velocity in phase space is zero.  The third term in the right-hand side of Eq. $\ref{continuity}$, with $\vec\nabla_{p} \cdot (\vec{F})$, gives us restrictions on the forces that can be applied to the particles. 
For this term to go to zero, we need the forces to be conservative and differentiable. In particular, they must be p-divergence free. Evidently, if collisions are present, they will not fulfill these requirements. Now, we can analyze the case of magnetic forces and can express this term explicitly: 
\begin{equation}
\vec\nabla_{p} \cdot (\vec{F}) = q \vec B\cdot (\vec\nabla_{p} \times \vec v) - q \vec v\cdot (\vec\nabla_{p} \times \vec B) 
\end{equation}

Assuming that the magnetic field is independent of the velocity of the particle, the p-curl of $\vec B$ goes to zero. In this case, the assumption is valid since the particles are moving so fast that the possibility of changing the magnetic field is negligible.  For the p-curl of $\vec v$, it just vanishes, since the velocity is the gradient of a scalar function, the relativistic energy. 
Therefore, if we have only pure magnetic forces, the $\vec\nabla_{p} \cdot (\vec{F})$ cancels. 

We arrive at the equation: 
\begin{equation}
\label{vlasov}
\frac{\partial \rho }{\partial t} + \vec{v}  \cdot \vec\nabla ( \rho ) + \vec{F}  \cdot  \vec\nabla_{p}( \rho) = 0
\end{equation}

But this is just the expression for the total derivative of the distribution. Therefore, we can reexpress it as:
\begin{equation}
\frac{d\rho}{dt} = 0
\end{equation}
which is the precisely the expression for Liouville's theorem ~\citep{goldstein_2002, bradt_2008b}.
\begin{figure*}[t!]
\vspace*{2mm}
\begin{center}
\includegraphics[width=\columnwidth]{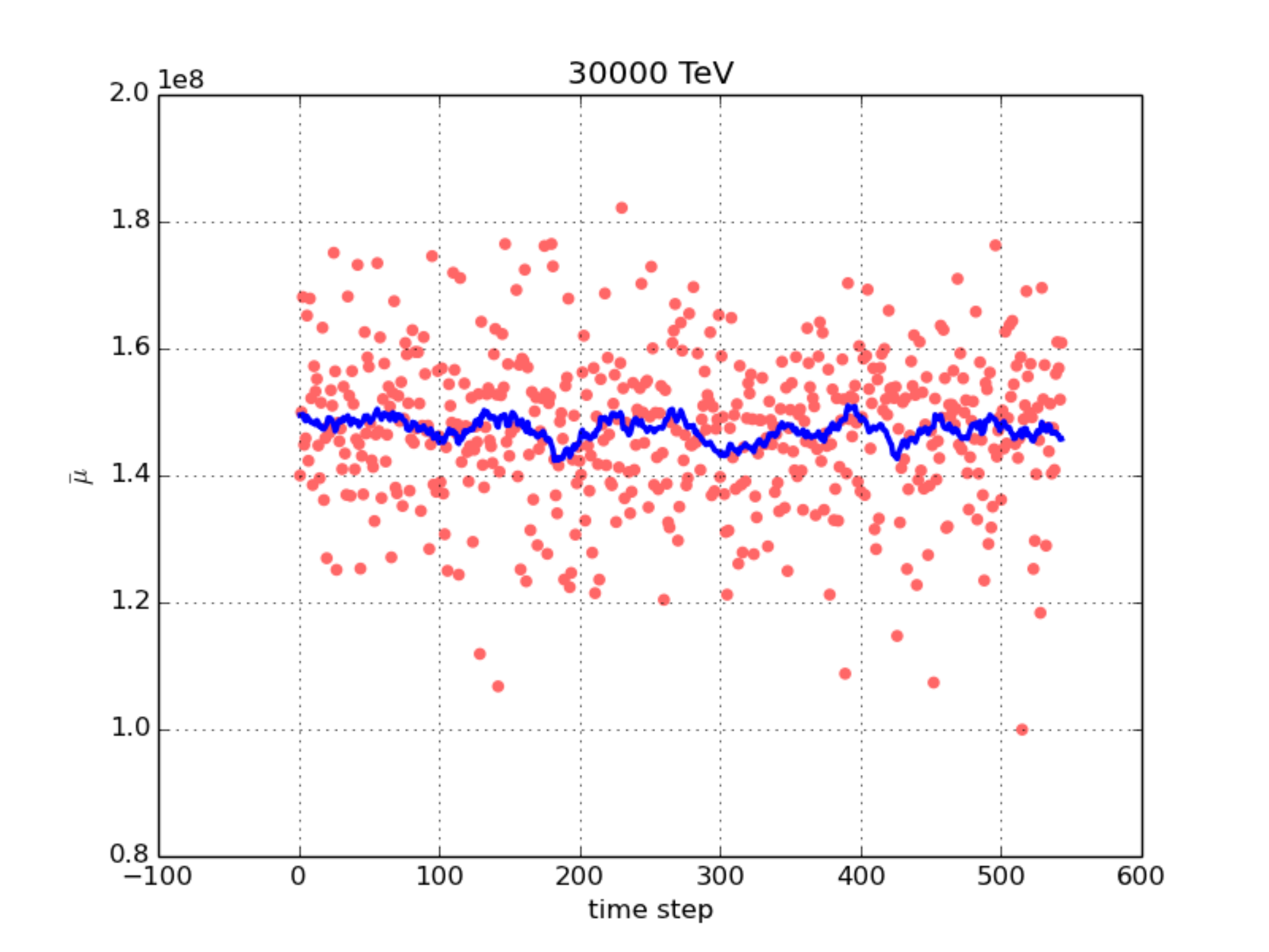}
\includegraphics[width=\columnwidth]{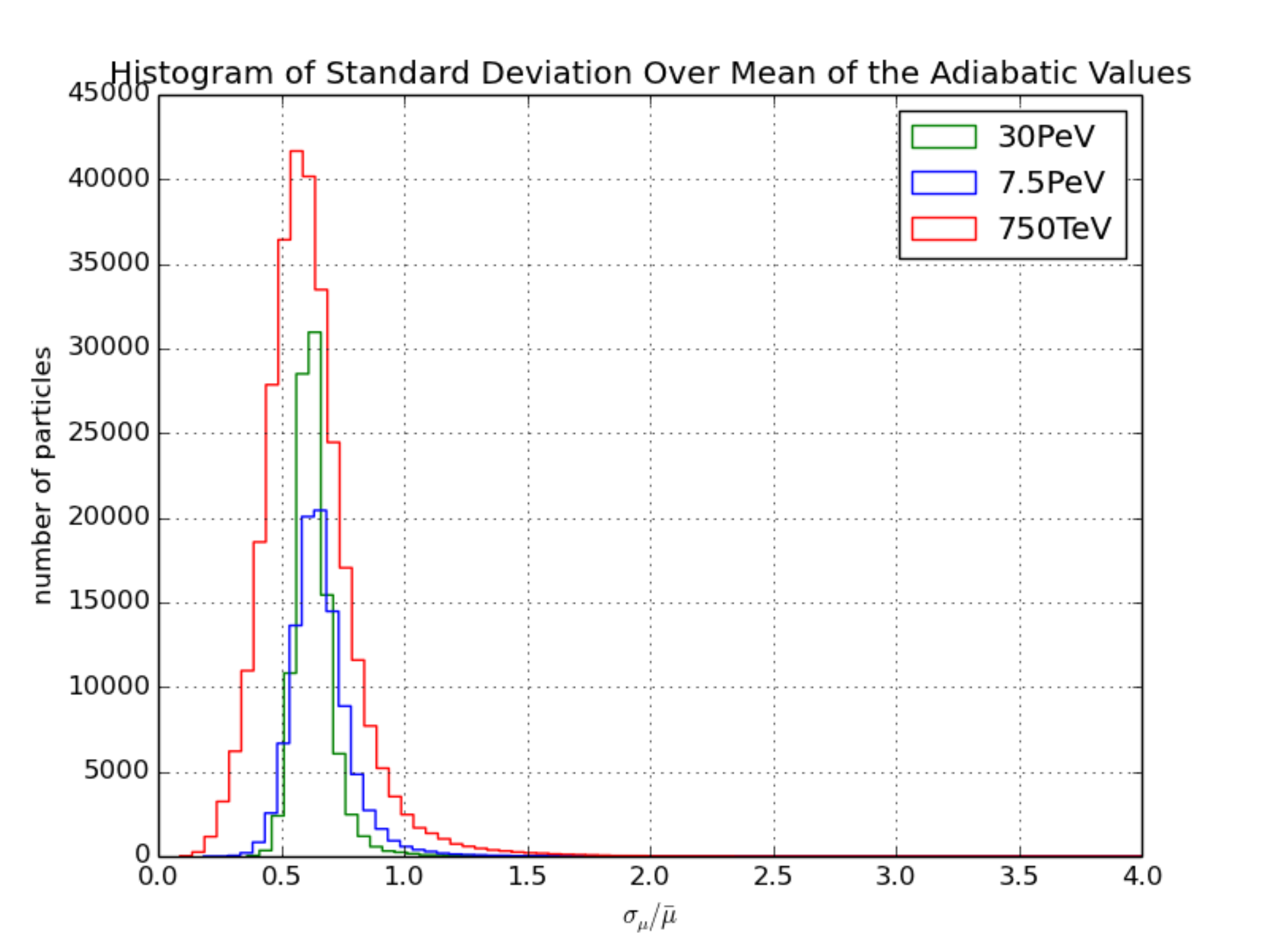}
\end{center}
\caption[Liouville conservation.]
{{\it Left}: The mean magnetic moment $\bar{\mu}$ for the data set at 30 PeV (red circles) with the corresponding moving average with subset of 30 time steps (blue line). The magnetic moment is calculated for all the particles at each time step, and then the average for the total set is calculated at each step. {\it Right}: Histogram of standard deviation of magnetic moment $\sigma_{\mu}$ over mean magnetic moment $\bar{\mu}$ for the three data sets of Table~\ref{tab:sets}. The red, blue, and green lines represent 750 TeV,  7.5 PeV, and 30 PeV protons, respectively. The magnetic moment is calculated for each particle at all time steps. The mean value and the standard deviation are for the total trajectory. Note that the 750 TeV set has three times the number of particles as the other sets, as described in Table $\ref{tab:sets}$.}
\label{fig:liouville}
\end{figure*}

This derivation is for a pure magnetic force, but in fact when calculating the particles' trajectories in a turbulent magnetic field, a variety of factors come into play. The most significant effect is when particles encounter a region where the magnetic field varies abruptly, i.e., the scale of variation of the magnetic field is shorter than the gyroradius of the particle. In this scenario, the trajectory does not have time to adjust smoothly to this change, and the interaction can be effectively considered a collision. For such cases, the right-hand side of Eq. $\ref{vlasov}$ can be modified by the addition of a term,
\begin{math} {\left [ \frac{\partial\rho }{\partial t} \right ]}_c \end{math}, which takes into account collisions of various origins that are differentiated by their exact functional form, given the fact that they will cause a nonzero time rate of change in 
the distribution function~\citep{baumjohann_1996}.
Therefore, under these conditions, Liouville's theorem can't be applied. To ensure that the abruptness in the trajectory is limited, we can calculate how adiabatic this change is. Having established this, it will ensure that a smooth variation will not greatly modify the density in phase space. 






In the presence of collisions, the magnetic moment of the gyrating particles changes. Therefore, to check for the adiabaticity of the trajectories, we can calculate the magnetic moment for each particle at each time step and find out if it statistically behaves as an adiabatic invariant.

The relativistic magnetic moment (also called {\it first} adiabatic invariant) is given by:
\begin{equation}
\mu = \frac{{p_{\bot}}^{2}}{2 m |\vec{B}|}
\label{eq:magmom}
\end{equation}
where $p_{\bot}$ is the momentum perpendicular to the magnetic field $\vec{B}$ and $m$ the particle mass. This quantity, relating magnetic field and perpendicular momentum of the particle, is conserved if the field gradients are small within distances comparable to the particle gyroradius. 
%
Using the integrated particle trajectories from the data sets of Table~\ref{tab:sets}, the magnetic moment Eq.~\ref{eq:magmom} was calculated at each integration time step and histogrammed in time step slices. In each time slice, the mean value $\bar{\mu}$ of the magnetic moment of the particle ensemble from each data set and the corresponding standard deviation $\sigma_{\mu}$ were calculated. Figure~\ref{fig:liouville} shows the evolution of $\bar{\mu}$ over the integration time of the 30 PeV set (on the left) of Table~\ref{tab:sets} and the ratio $\sigma_{\mu}$/$\bar{\mu}$ for the three sets (on the right) of Table~\ref{tab:sets}.
A perfect distribution with $\sigma_{\mu}$/$\bar{\mu}$ = 0, indicates that the particles most likely stay in the same magnetic field line, which is unrealistic for a particle moving along a turbulent magnetic field. Nonetheless, a distribution peaked at a value much larger than one will tell us that the particles suffered strong variations in their trajectories and collision-like interactions happened.
The distributions calculated for our three different energy data sets peak at around 0.5 and do not appear to have trends or large variations during integration time (compared to $\sigma_{\mu}$), as shown on the left of Figure~\ref{fig:liouville}. This indicates that even though the particles have interacted with the turbulent field, and it has changed their trajectories, the changes are relatively smooth and with limited statistical impact on the overall particle ensemble. 

Note that the width of the distributions in $\sigma_{\mu}$/$\bar{\mu}$ means that at some level the magnetic moment of the particles fluctuates during propagation. Different particles follow different magnetic field lines and experience different interactions. There might be also a contribution from the limited accuracy of the numerical integration. However, as discussed and shown in appendix~\ref{app:a}, in the present application, the effects due to accuracy limitations are not sufficiently large to significantly violate adiabaticity.
The highest level of possible inaccuracy, where particles are stochastically re-distributed at each gyration, provides a numerical diffusion at the level of Bohm diffusion, where particles are scattered every gyration. It has been proven that in astrophysical turbulence Bohm diffusion is always much smaller than diffusion induced by magnetic field line wandering~\citep{ly_2014}. So the fact that the accuracy of the used trajectory integration method is significantly below the Bohm diffusion level (see appendix~\ref{app:a}) poses no problem on the statistical results obtained in this study.

Based on the fact that the ensemble-average $\langle \sigma_{\mu}$/$\bar{\mu}\rangle \lesssim$ 1, especially that only magnetic forces are explicitly taken into account and that the changes in the particles' trajectories are relatively smooth, it is assumed that Liouville's theorem is applicable (see further discussions in Section~\ref{sec:disc}). So in this study, back-propagation of the particle trajectories is justified in this statistical sense. The situation changes if scattering rate is higher, such as in the heliospheric magnetic field case studied in~\cite{barquero_2016}, where resonant scattering processes produce larger deviations from adiabaticity.


%

\section{Results}
\label{sec:res}

This section shows the results obtained with the numerical calculation data sets described in Section~\ref{ssec:crprop}.

\subsection{Mean Free Path}
\label{ssec:mfp}



%
\begin{figure*}[t!]
\vspace*{2mm}
\begin{center}
\includegraphics[width=\columnwidth]{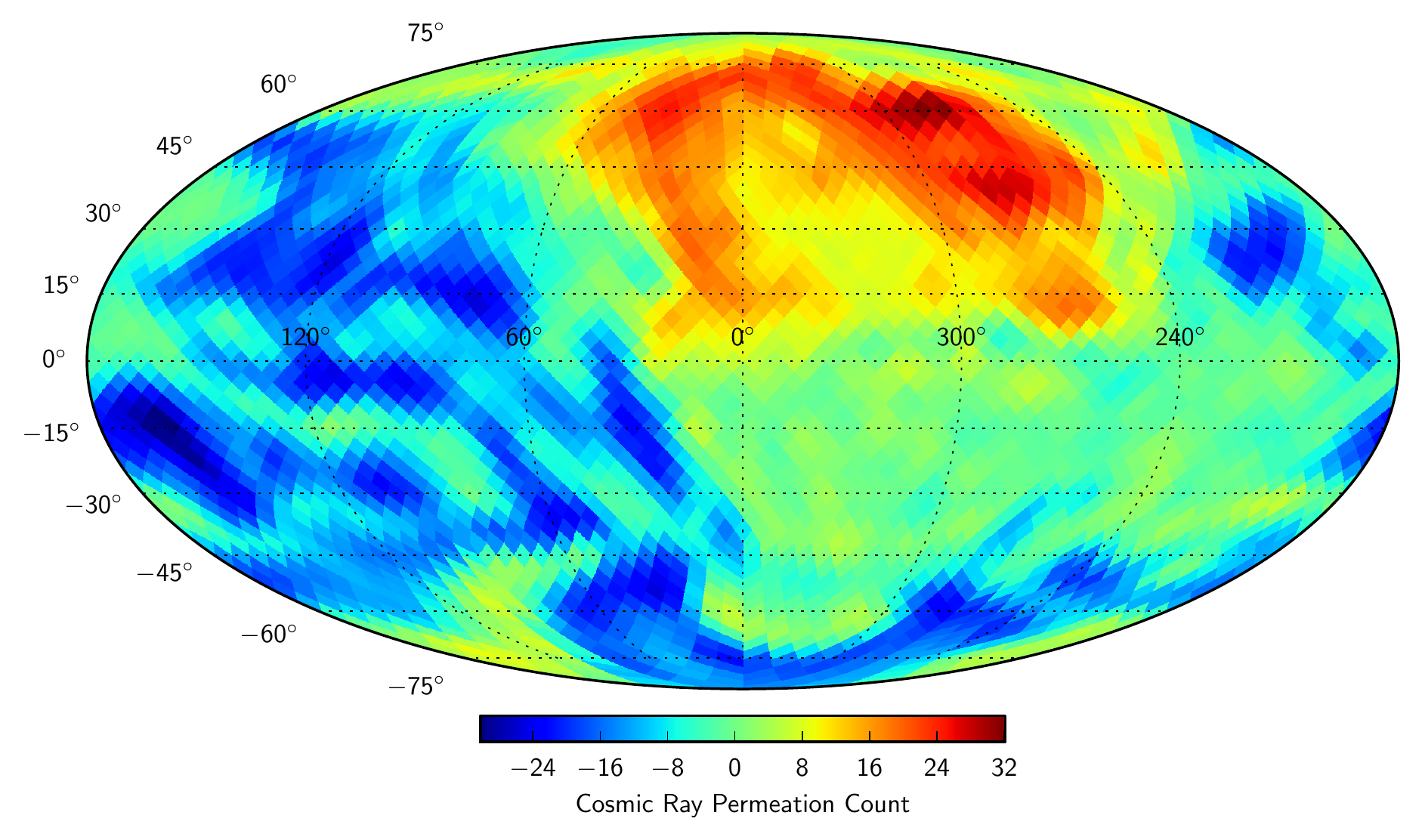}
\includegraphics[width=\columnwidth]{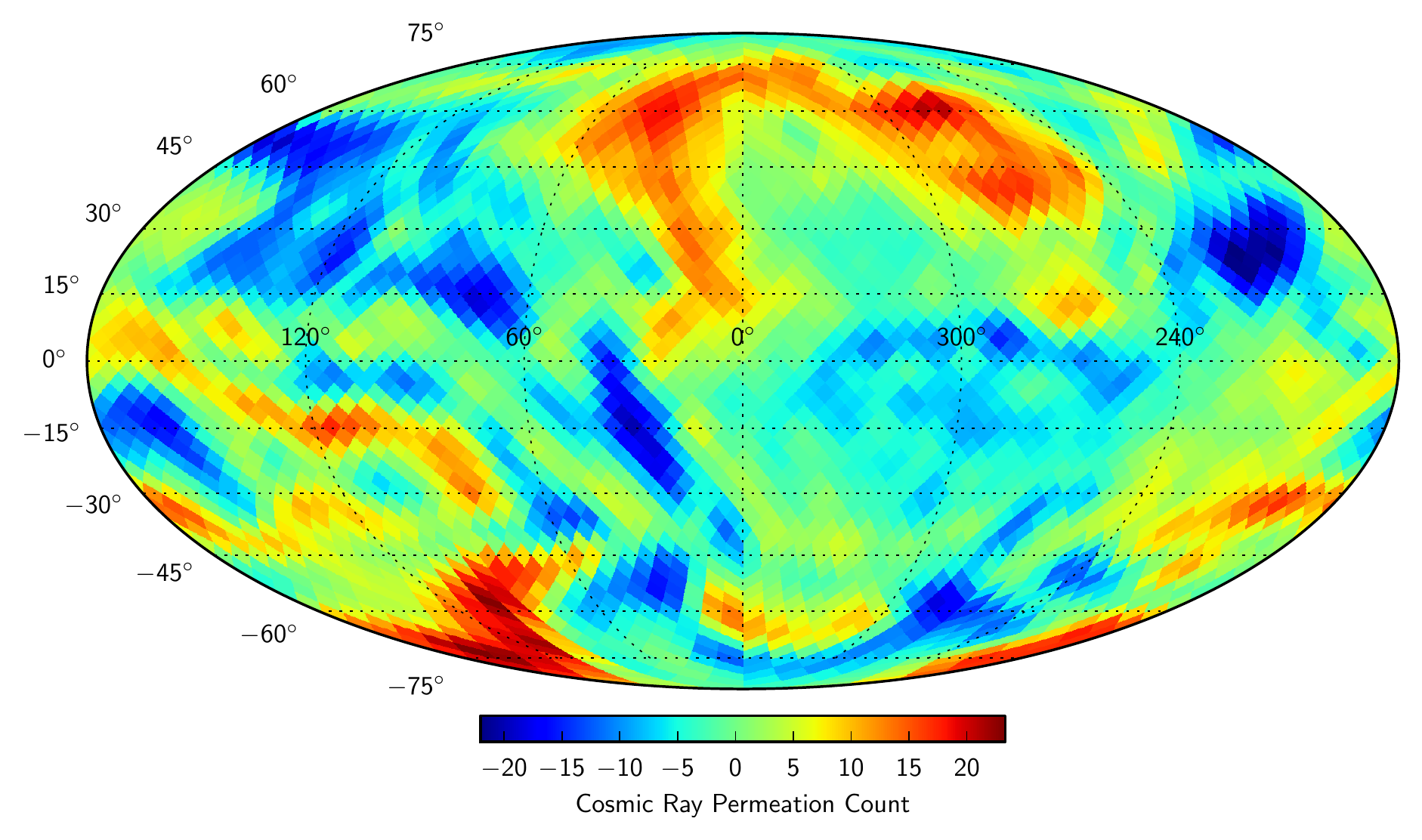}
\end{center}
\caption{
Sky map of arrival direction distribution of the 7.5 PeV proton set of Table~\ref{tab:sets} after propagation for a distance of 40 pc. The decomposition of the initial dipole distribution is shown. On the left is the sky map obtained after time inversion, and on the right is the same map after subtracting the dipole component from the map on the left. A Gaussian smoothing with $\sigma = 3^{\circ}$ was used.}
\label{fig:dipremoval}
\end{figure*}
%
\begin{figure}[h!]
\flushleft
\includegraphics[width=\columnwidth]{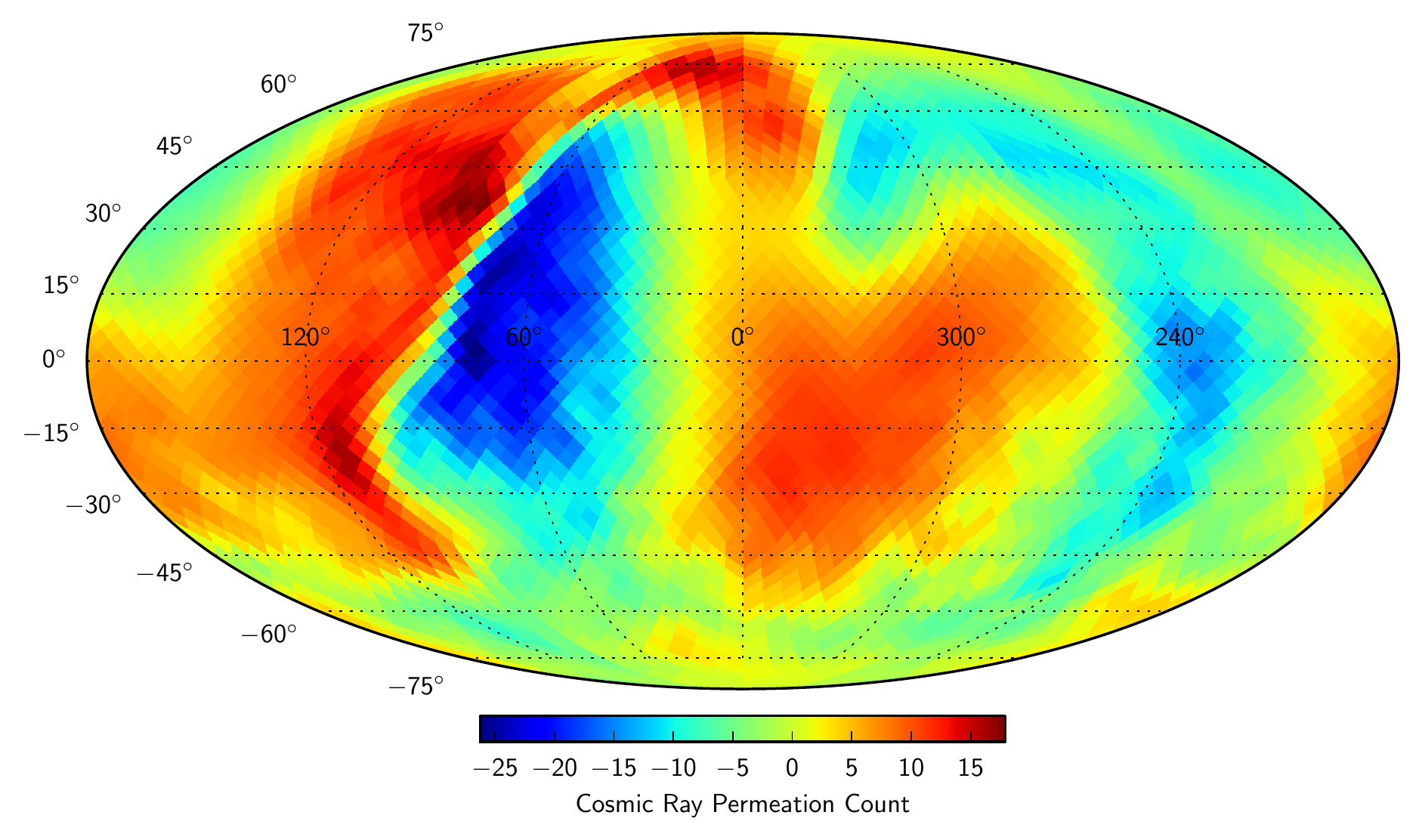}\\
\includegraphics[width=\columnwidth]{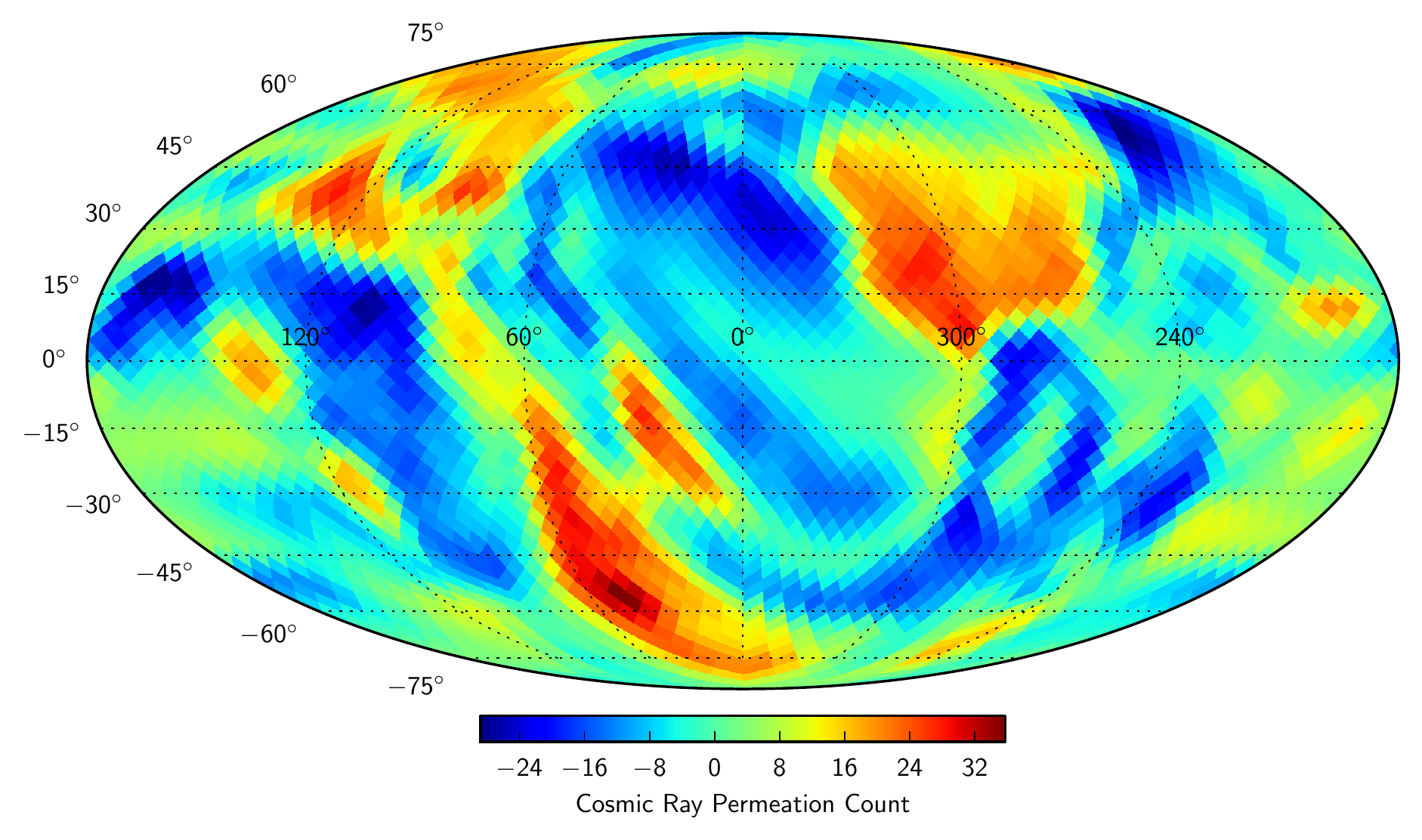}\\
\includegraphics[width=\columnwidth]{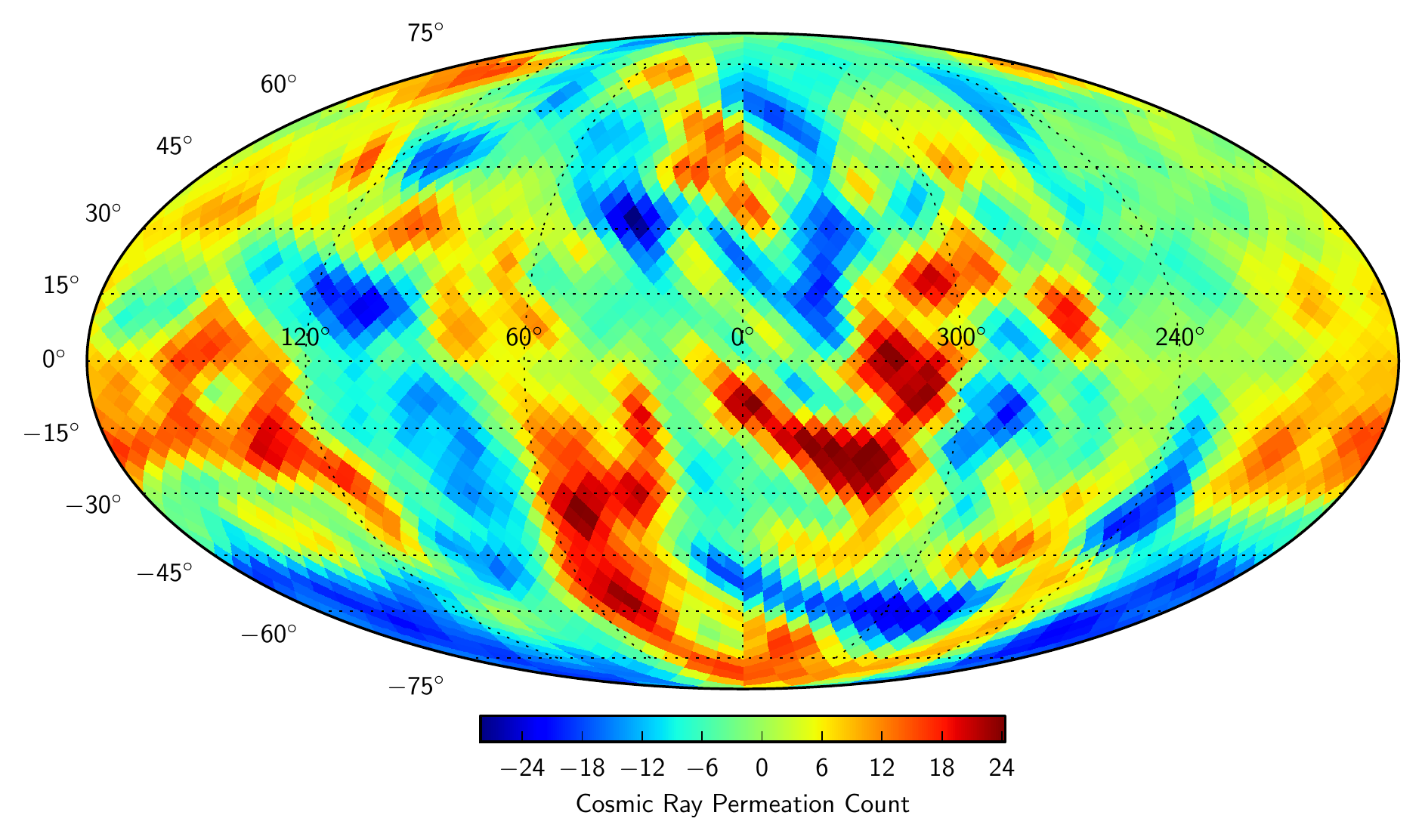}\\
\includegraphics[width=\columnwidth]{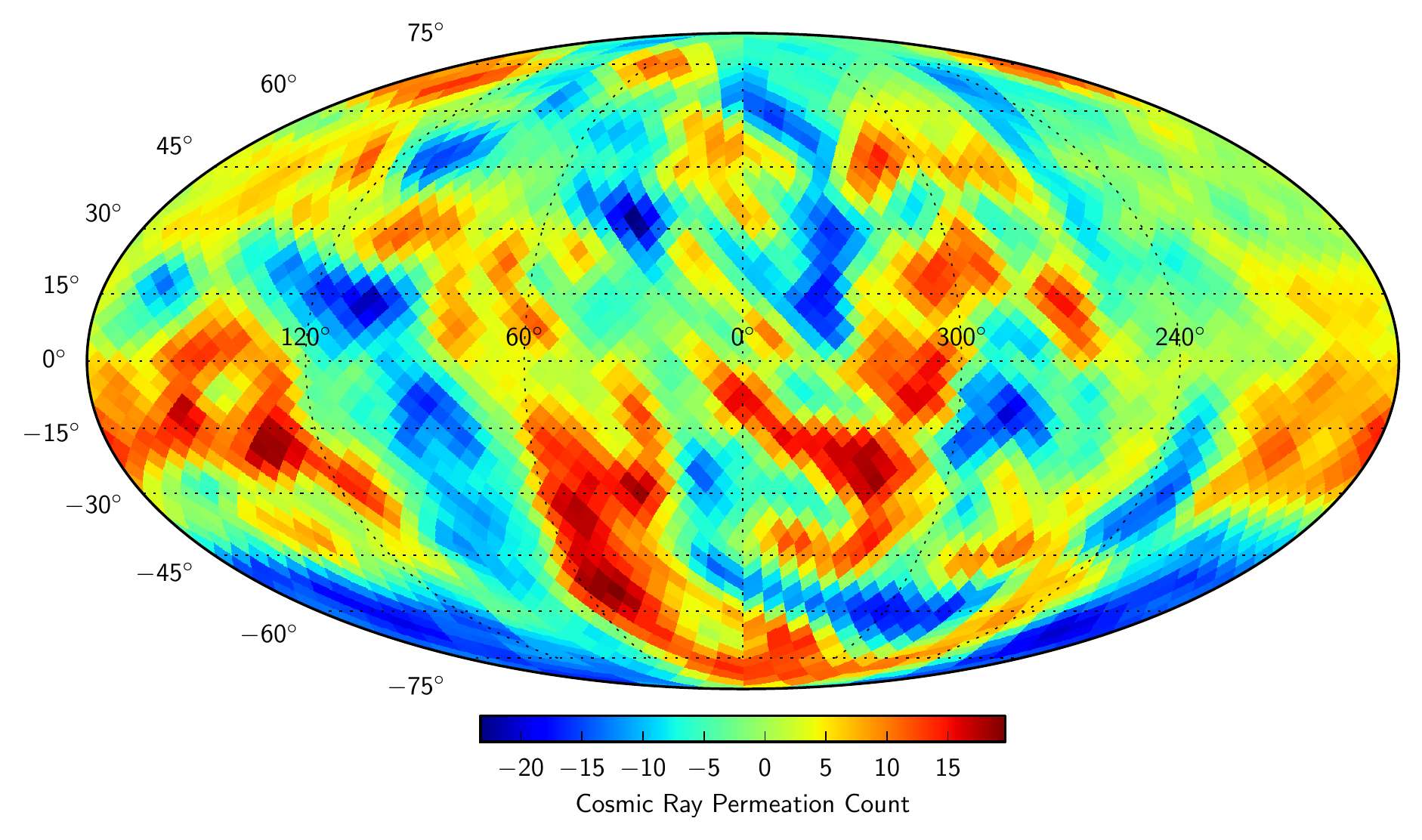}
\caption{
Sky maps of arrival direction distributions of 30 PeV protons in equatorial coordinates, with the dipole density gradient weight at different distances: $R$ = 10 pc, 20 pc, 60 pc, and 90 pc (from top to bottom). Gaussian smoothing with $\sigma = 3^{\circ}$ was used. On each map, a dipole fit was performed and the resulting dipole component was subtracted.}
\label{fig:progression}
\end{figure}

Using the particle trajectory data sets from Table~\ref{tab:sets}, it is possible to evaluate general properties that the ensemble of particles have after a sufficiently long duration of propagating in the turbulent magnetic field. The mean free path $\lambda_{mfp}$ is the distance
at which the instantaneous particle directions become uncorrelated with respect to those at time t=0. At this distance, and associated to scattering time scale, particles have lost {\it memory} of the direction distribution at initial conditions. This is a cumulative property of all particles, and it can be estimated by calculating the mean distance at which the direction of each particle has an angle of 90$^{\circ}$ from its initial position. This definition is equivalent to evaluating the velocity autocorrelation function and estimating the distance at which it is sufficiently close to zero (i.e., correlation to initial condition is lost).
%

The results of $\lambda_{mfp}$ calculations are given in Table $\ref{tab:sets}$. As expected in this regime, the mean free path increases with energy. For the interarm region, where the injection scale is on the order of 100 pc, $\lambda_{mfp}\sim$ 60 pc for 30 PeV protons. For an energy four times smaller, 7.5 PeV, the mean free path decreases to a value of 50 pc. In the spiral arm, with injection scale on the order 10 pc, our calculation for 750 TeV protons is 5.0 pc.
Note that in the two lowest energy data sets considered here, the particle gyroradius is on the order of the damping scale of the turbulent field.
Therefore, pitch angle scattering arising from smaller scale magnetic perturbations is significantly reduced, and may result in the mean free path being overestimated to some degree. In the present work, the intent is to show the effects of turbulence in the particle spatial distribution in relation to the mean free path; therefore, whether the value is strictly correct is of limited importance in this context. In future studies, the need to use MHD simulations in a wider inertial range will be considered in more detail.

\subsection{Sky Maps of Arrival Direction Distribution}
\label{sec:skymaps}
%
\begin{figure*}[t!]
\begin{center}
\includegraphics[width=\columnwidth]{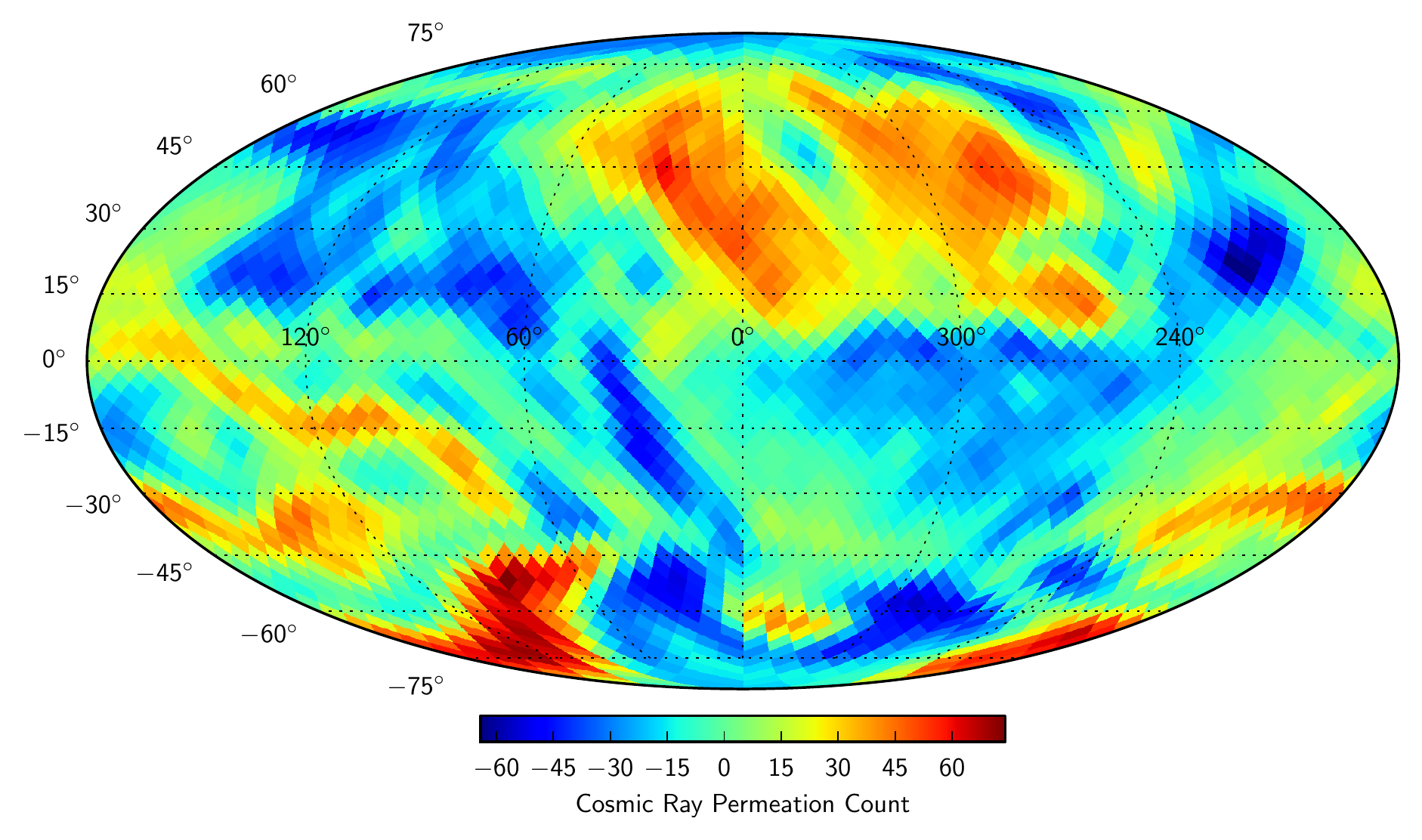}
\includegraphics[width=\columnwidth]{30PeV_smooth_New_Map_at_Earth_Dipole_Removed_Radius_60pc_dip_is1.pdf}
\end{center}
\caption{
Sky maps of arrival direction distributions of 750 TeV protons (on the left) and 30 PeV (on the right) in equatorial coordinates and at propagation distance corresponding to the mean free path. Gaussian smoothing with $\sigma = 3^{\circ}$ was used. On each map, a dipole fit was performed and the resulting dipole component was subtracted.}
\label{fig:mapcomparison}
\end{figure*}

%
The particle trajectories numerically integrated with Eqs. 1-2 for the sets in Table~\ref{tab:sets} are used to study the properties of their arrival direction distribution that results from the scattering off magnetic turbulence from a particle density gradient.
As described in Section~\ref{ssec:crprop}, the procedure used for determining the sky maps of the particles' arrival distribution makes use of the trajectory integration backward in time, uniformly emanating from one point assumed to be Earth, until they exit the integration box. At a sufficiently long distance from the origin, particle trajectories accumulate in the direction of the mean magnetic field, since the perpendicular diffusion is smaller than that which is parallel to the magnetic field lines. A sphere of radius $R$ with center at the origin of the back-propagated trajectories is used to record the position and velocity direction of each trajectory. The trajectories are inverted in time from those locations on the sphere to the origin, by virtue of Liouville's theorem (see Section~\ref{sec:lt} for a discussion of the validity of Liouville's Theorem and the consequences of its use in the context of the present study).

It is likely that there are several sources of cosmic rays in the Milky Way, perhaps from different populations injecting particles into the ISM with different energy spectral shapes. It is also possible to suppose that a single source may dominate the cosmic ray distribution at Earth, depending on the energy range (see Section~\ref{sec:intro}). In a scenario of isotropic diffusion, the cosmic ray density gradient is expected to be described by a simple dipole distribution. This is a similar distribution as would be expected if convective transport were the dominant source of the cosmic ray density gradient, such as in the case of superbubbles~\citep{biermann_2013} or the effect of Loop I shell in the local ISM~\citep{schwadron_2014}. In the general and more likely scenario of anisotropic diffusion, cosmic ray propagation along the magnetic field line is faster than the perpendicular, thus producing a cosmic ray gradient ordered along the regular magnetic field (see, e.g.,~\cite{kumar_2014,mertsch_2015}). Although the density gradient is not expected to be a dipole but rather a distribution that depends on the ratio of perpendicular to parallel diffusion, in this work it is assumed a simple dipole distribution, regardless of the origin of the underlying density gradient of the cosmic rays.

If the underlying distribution of cosmic rays is perfectly uniform, the effects of scattering off magnetic turbulence shuffles one isotropic distribution into another isotropic distribution. However, with an existing particle density gradient, the scattering processes redistribute particles from the region of higher density to that of lower density, and vice versa, thus creating a complex arrival distribution that can be described with higher order multipole terms of the spherical harmonic expansion.
%

In the process of inverting time on the numerically calculated trajectories, a dipole gradient distribution, assumed to be aligned with the direction of the mean magnetic field of the MHD snapshot, is introduced as a weight on each trajectory at the crossing point on a sphere of radius $R$. The weight is calculated using the angle of the particle direction at the crossing point from that of the density gradient. The arrival distribution of these forward-propagated trajectories at Earth (i.e., the origin) is then determined.
One key factor to remember is that only the small-scale angular anisotropy is studied, since this is the one that arises from the specific interaction with the turbulent magnetic field. Therefore, the large scale component, specifically the dipolar component of the map at Earth, is fitted and removed. Such a dipole component can have a different direction and amplitude than that injected. Figure~\ref{fig:dipremoval} shows the sky map, in equatorial coordinates, of the arrival distribution of the 7.5 PeV protons at Earth before (on the left) and after (on the right) dipole subtraction, thus emphasizing the small scale features. A Gaussian smoothing with $\sigma = 3^{\circ}$ was used. The residual map shows medium- and small-scale angular structures arising from the breakout of the underlying dipole anisotropy after a propagation of $R$ = 50 pc.

The maps were created with the HealPix mapping tool~\citep{gorski_2005}, which divides the sphere into pixels of equal areas. For the present work, a pixelization parameter of $N_{side}=16$ was used, which corresponds to 3072 pixels in total, with a mean spacing of $3.67^{\circ}$. In Figure~\ref{fig:dipremoval}, the excess regions, with respect to the average isotropic flux, are identified by a red color and deficit regions by a blue one. Therefore, a pixel in which many particles pass through will be represented in a stronger red color than one that has only a few events. 

Figure~\ref{fig:progression}, under the assumption of the Earth's location in the interarm zone with $L_{inj}$=100 pc, shows the sky maps progression of the 30 PeV cosmic ray arrival direction distribution with the dipole density gradient weight at different propagation distances of $R$ = 10 pc, 20 pc, 60 pc, and 90 pc. On each map, a dipole fit was performed, and the resulting dipole component was subtracted. Such a component may be different for each map, since it depends on the actual magnetic field structure at a given distance. The sky maps show that by increasing the propagation distance the arrival direction distribution progressively develops smaller structures up the mean free path (60 pc in this case, see Table~\ref{tab:sets}). At larger distances, the arrival direction distribution reaches a statistically steady configuration (in Figure~\ref{fig:dipremoval} only the 90 pc propagation distance is shown). Only propagation processes within the mean free path are responsible for the actual arrival direction distribution of cosmic rays at Earth, as discussed also in~\cite{giacinti_sigl_2012}. Whatever happens at larger distances is reshuffled by the scattering processes and is only relevant in the generation of the seed particle density gradient at large scale.

The actual distribution of cosmic rays depends on the specific realization of the turbulent magnetic field and on the particle energy. Figure~\ref{fig:mapcomparison} shows the steady-state arrival direction distribution of 750 TeV (on the left) and 30 PeV (on the right) cosmic rays. This figure qualitatively shows that in the higher energy case the anisotropy of the distribution shows more small-scale angular regions than in the lower energy case. For a quantification of such a visual property, it is necessary to calculate the angular power spectrum.
The map from 7.5 PeV presents almost the same distribution as the one for 750 TeV, since the only differences between them are the assumption on the injection scale and that they are independent sets (as described in Section~\ref{ssec:crprop} and shown in Table~\ref{tab:sets}). The particles at 7.5 PeV energy and $L_{inj}=100 pc$ are physically equivalent to particles at 750 TeV with $L_{inj}=10 pc$.

\subsection{Angular Power Spectrum}
\label{sec:aps}
%
\begin{figure*}[t!]
\begin{center}
\includegraphics[width=1.5\columnwidth]{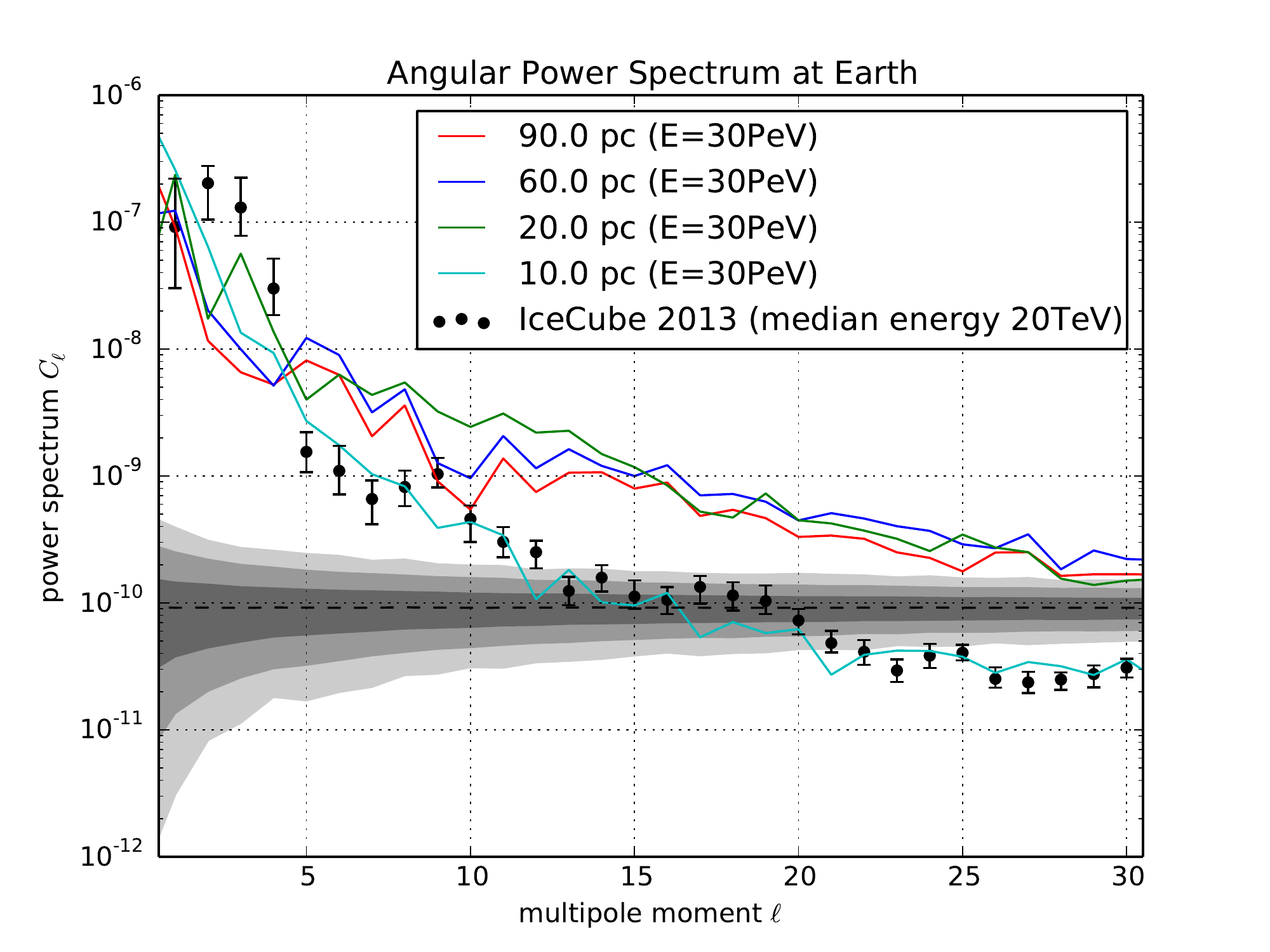}
\end{center}
\caption{Angular power spectrum of the arrival direction distribution of 30 PeV trajectories set of Table~\ref{tab:sets} and Figure~\ref{fig:progression} with dipole weight injected at a distance of 10 pc (in cyan), 20 pc (in green), 60 pc (in blue and corresponding to the mean free path) and of 90 pc (in red). The gray bands show the 1$\sigma$, 2$\sigma$ and 3$\sigma$ bands for a large set of isotropic sky maps. The black circles are the results from the IceCube observatory at a median energy of 20 TeV~\citep{santander_2013}. Note the difference in energy scale between the experimental data and the numerical calculations.}
\label{fig:aps30PeV}
\end{figure*}
The sky maps of arrival direction distributions described in the previous sections are not dissimilar to experimental observations. However, it is the determination of the angular power spectrum that makes it possible to quantify the formation of small-scale structure anisotropy arising from scattering off magnetic turbulence. 
With the power spectrum, a sky map of arrival direction distribution is decomposed into spherical harmonics to provide information on the angular scale contribution of the anisotropy. The spectrum quantitatively indicates which multipole moments $\ell$ in the spherical harmonic expansion contribute to the observed map. The IceCube observatory provided a power spectrum of their Southern Hemisphere observation in the 10 TeV scale~\citep{abbasi_2011,aartsen_2015} and the HAWC gamma-ray observatory provided one for the Northern Hemisphere in the TeV scale~\citep{abeysekara_2014}.
The angular power spectrum is determined using the {\tt anafast} tool in the HealPix framework.

Figure $\ref{fig:aps30PeV}$ shows the angular power spectrum from the numerical trajectory integration set of 30 PeV protons at propagation distances from the initial unperturbed dipole anisotropy at 10 pc, 20 pc, 60 pc and 90 pc, as in Figure~\ref{fig:progression}, but without subtracting the dipole component. In the figure, the result from the observations made by the IceCube observatory is included as well. Note that the higher power values observed at low $\ell$, not reproduced in the calculations, are most probably due to the partial sky view of the IceCube observatory, which causes correlations with the largest scale spherical harmonic moments. The experimental observation is for a median cosmic ray energy of about 20 TeV, much lower than the numerical calculations. The numerical calculation sets are normalized to the dipole component (i.e., $\ell$ = 1) of experimental power spectrum for the farthest propagation distance of 90 pc only. At shorter propagation distances, the normalization corresponds to the relative power obtained from the calculations. This normalization is valid since we are interested in the small-scale features, not on the assumptions on the large-scale anisotropy.
\begin{figure*}[t!]
\begin{center}
\includegraphics[width=\columnwidth]{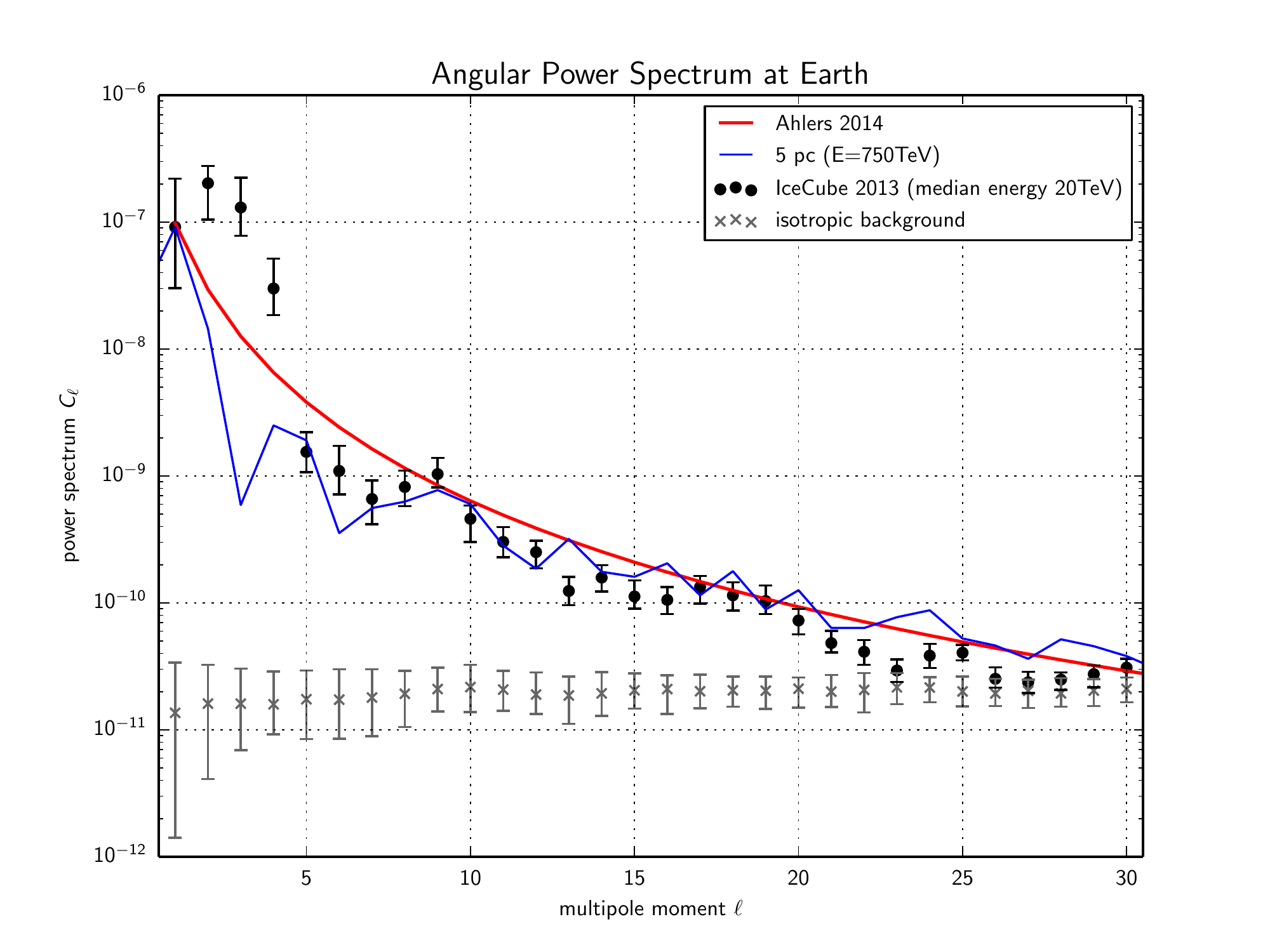}
\includegraphics[width=\columnwidth]{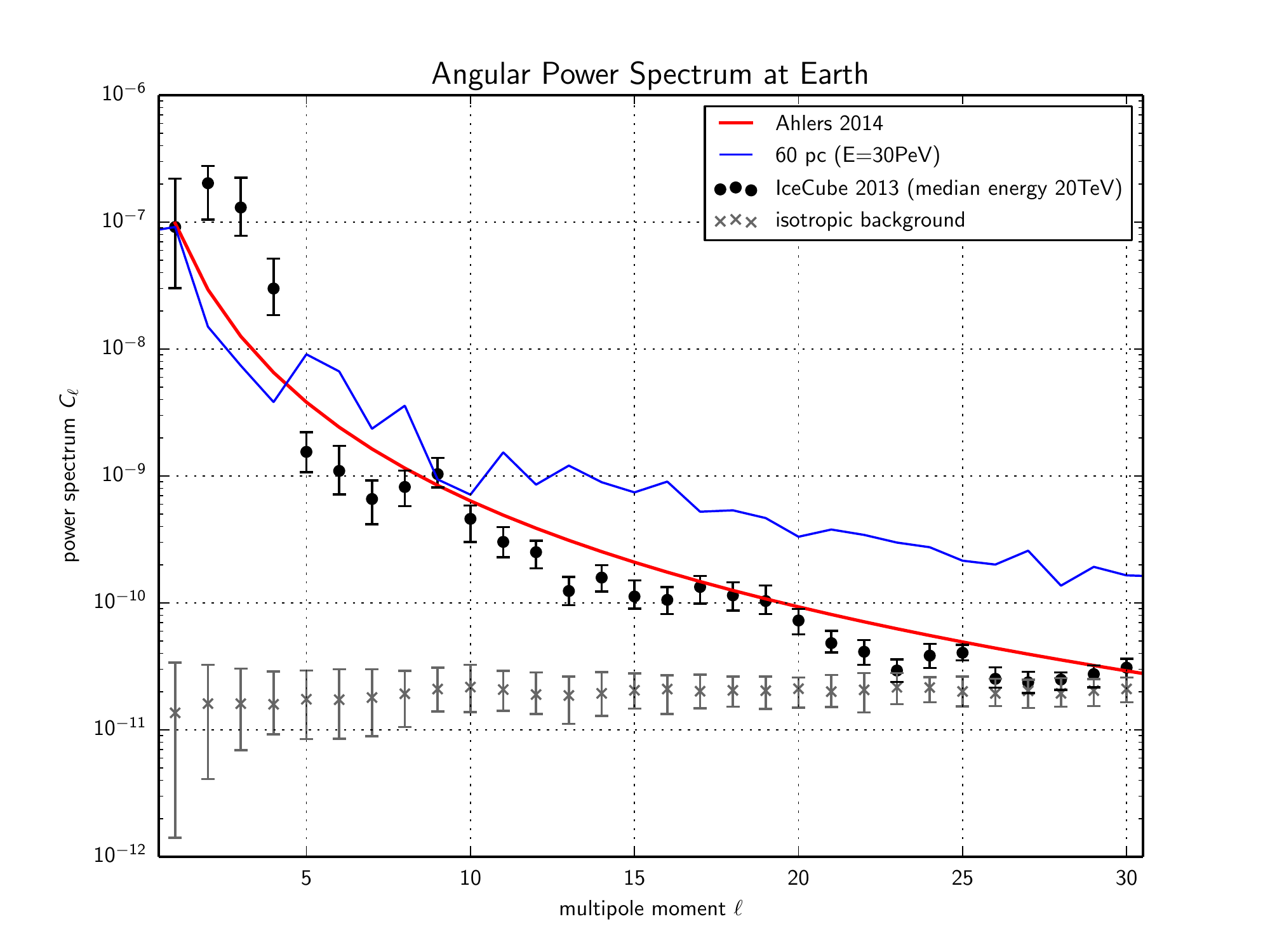}
\end{center}
\caption{Angular power spectrum of the arrival direction distribution of 750 TeV (blue line on the left) and 30 PeV (blue line on the right) trajectories sets of Table~\ref{tab:sets} with dipole weight injected at the corresponding mean free path distance. The red line is the power spectrum from~\citep{ahlers_2014}. The black circles are the results from the IceCube observatory at a median energy of 20 TeV. The gray crosses and error bars show the 1$\sigma$ band for a large set of isotropic sky maps~\citep{santander_2013}. Note the difference in energy scale between the experimental data and the numerical calculations.}
\label{fig:apscomparison}
\end{figure*}

Note that the numerical calculation shows, within statistical fluctuations, that the power of the dipole component decreases with increasing propagation distance, due to the transfer of part of it into the higher $\ell$ components.
At very short distances, i.e., smaller than the mean free path, the low multipole moments are dominant. However, as the distance increases, more power is transferred to the higher multipole moments, which is the signature of particle interactions with the turbulent magnetic field. Once they reach the mean free path distance, in this case 60 pc, the power spectrum converges to a steady configuration, as the sky maps in Figure~\ref{fig:progression} show.
In Figure~\ref{fig:aps30PeV}, the power spectrum of an isotropic flux of the same number of particles is included as a gray band and properly normalized~\footnote{The power spectrum of the isotropic flux is calculated by generating 10,000 realizations of uniform particle direction distribution, matching the number of the integrated particle trajectories, and calculating the power spectrum for each of them. The 68\%, 90\%, and 99\% containment bands are reported in Figure~\ref{fig:aps30PeV}}.
It is evident that the distribution at distance $\lambda_{mfp}$ and beyond is not only a random distribution but also possesses a definite structure. It is noted that the angular power spectra calculated for propagation distance longer that 90 pc show the same relative normalization and shape of that corresponding to the mean free path. This is compatible with the achievement of steady state in the anisotropy structures as evident in the sky maps of Figure~\ref{fig:progression}. The approach of the power spectra to the band of isotropic distribution for $\ell >$ 30 means that, for the numerical realization studied here, the smaller angular scales are indistinguishable from random fluctuations of the isotropic flux.

As observed in Figure~\ref{fig:mapcomparison}, the angular structure generated from the effect of scattering off magnetic turbulence depends on the particle energy (750 TeV and 30 PeV shown). This is evident also in the corresponding power angular spectra in Figure~\ref{fig:apscomparison}, calculated at the propagation distance of their mean free paths.
The figure also shows the experimental result from the IceCube observatory at 20 TeV median energy, with the corresponding power of the isotropic distribution background and with the curve expected from the hierarchical breakup of angular components from~\cite{ahlers_2014}, normalized to the dipole component of the IceCube result. Note that the angular power spectrum at higher energy is flatter than that at lower energy. This is compatible with the existence of more small-scale structures as evident in the sky maps of Figure~\ref{fig:mapcomparison}.



\section{Discussion}
\label{sec:disc}

We have shown how small-scale anisotropy arises from the interaction of particles with the turbulent magnetic field. Specifically, we have addressed how the integration of trajectories in an MHD turbulent magnetic field provides a realistic understanding of the small-scale features present in the observations of cosmic ray anisotropy at Earth. The ISM is in a plasma state, where the MHD equations serve as a model for its dynamics, therefore an MHD turbulent magnetic field is a natural approach to study the magnetic field properties in the ISM. 
Previous work~\citep{giacinti_sigl_2012} had considered the effects of magnetic turbulence on cosmic ray distribution. In this study, the authors have considered synthetic turbulence, which, on one hand, lacks the proper development of the gas dynamics but, on the other hand, provides the first qualitative connection between a magnetic turbulent field and cosmic ray arrival directions. 
%
In compressible MHD turbulence, scattering efficiency strongly depends on the wave type and how the particle gyroradius compares to the turbulence scales. Specifically, fast modes are identified as the main source for cosmic ray scattering~\citep{yan_lazarian_2008}.

The dynamics of the different turbulence modes and the relationship between particle energy and turbulence scale determine the actual scattering efficiency, which is most responsible for the breakout of a global cosmic ray density gradient into small-scale angular anisotropy. The angular power spectrum calculated for the data sets studied in this work appears to evolve in time until particles cross the mean free path distance. In a steady state, the shape of the angular power spectrum is a function of the magnetic field structure and of the consequent effects on particle propagation.

Studying particle trajectories in MHD magnetic turbulence provides a more realistic framework to investigate the behavior of cosmic ray propagation in the ISM, where turbulence is expected to be anisotropic, although MHD turbulence simulations typically represent a significantly more restrictive inertial range than the actual astrophysical plasmas.

The ISM is a complex environment and its exact representation is difficult to achieve; therefore, our MHD magnetic field can be considered one possible configuration of the magnetic field in the ISM. For this reason, direct comparison should be on the angular power spectrum itself, not on the exact topology of the small-scale features in our maps.

The framework for the application of Liouville's theorem is provided in the context of cosmic ray arrival distribution and applied through the back-propagation method. Although particle trajectories appear to suffer from mild deviation from adiabaticity, Liouville's theorem was applied in this particular case to study the first order effects of magnetic turbulence on the global topology of particle trajectories. This is because particles do not experience severe scattering in their trajectories, as shown in the first adiabatic invariant calculations; nonetheless, if the magnetic field were to vary dramatically with respect to the gyroradius of the particles, it prohibit application of Liouville's theorem. 
%
The spread in the magnetic moment $\mu$ distribution in Figure~\ref{fig:liouville} suggests that some trajectories may have experienced more scattering than average in their propagation. This effect will manifest in the anisotropy through higher power at high $\ell$, since these particles will have had greater interactions with the turbulent field, resulting in a slightly flatter angular power spectrum. If the distribution on the $\sigma_{\mu}/ \bar{\mu}$ plot had peaked at a higher value or a progressive trend had appeared on the mean magnetic moment plot of the data set, this would be an indication of a clear violation of Liouville's theorem. Neither of these trends have statistically or significantly occurred in our calculations; nevertheless, we would expect them to appear in magnetic fields that have a strong variation in scale on the order of the gyroradius of the particles. 
Future studies will need to use MHD turbulence simulation with wider inertial range to include enlarge the energy range in which magnetic turbulence affects cosmic ray distribution. 
In this way, these studies will reveal the connection between the angular power spectrum of the cosmic ray arrival direction distribution and the turbulence properties, naturally accounting for spurious effects derived by the numerical methods used.
%

As mentioned in Section~\ref{ssec:crprop}, the trajectory back-propagation method is intended to provide a high efficiency in the studies of particle propagation in magnetic fields. Such a numerical method provides a mapping of the regions in space that are magnetically connected to the arrival point, where particles are assumed to be isotropically distributed. Therefore, appealing to the conservation of the total power across all spherical harmonic contributions, as dictated by Liouville's theorem, makes higher multipole moments possible. Anisotropy was studied as a function of an initial large-scale density gradient at a large distance from the arrival point. Such a density gradient, however, is generated by the same propagation processes that produce all angular structures in the arrival direction distribution as well. In fact, smaller scale structures can arise in trajectory forward-propagation integration methods, where particle escape is naturally accounted for, even without assuming an initial global anisotropy, as shown in~\cite{rettig_pohl_2015}. On the other hand, as long as a global anisotropy is developed at some distance larger than $\lambda_{mfp}$, independently on how it is generated, the arrival direction distribution reaches a steady state angular power spectrum, and the effects of $seed$ anisotropy and observed anisotropy can be disentangled.
%

For the mean free path calculation, it is shown to be dependent on energy. In this regime, from 750 TeV to 30 PeV, the $\lambda_{mfp}$ increases with energy. This is in agreement with how the more energetic particles interact with the perturbations of the magnetic field. In the case of the 750 TeV and 7.5 PeV sets, the minimum scale in the power spectrum of turbulence is of the size of the gyroradius of the particles, which may cause an overestimation of the $\lambda_{mfp}$, since there is limited power in the physical perturbations that interact with the particles. For our 30 PeV particles, we are well above this minimum scale, so the $\lambda_{mfp}$  is unaffected.




The anisotropy maps and their corresponding angular power spectrum can tell us about the propagation of the particles themselves in the turbulent field. At very short distances, the low multipole moments are completely dominant and hold most of the power, as shown with the line of 10 pc in Figure~\ref{fig:aps30PeV}. The reason for this is that the particles have not had enough time to interact with the features of the magnetic field, and the distribution is still reminiscent of the initial configuration.
However, the particles continue to interact and structures start to develop. As we can see with the line of 20 pc in Figure~\ref{fig:aps30PeV}, the higher multipole moments start to rise, with small-scale features becoming highly visible even in the maps in Figure~\ref{fig:mapcomparison}. 
One interesting feature is that the small-scale structures develop within the mean free path, but once they reache this scale, the maps do not change significantly. This observation is confirmed in the angular power spectrum. Therefore, the distribution of power between the different multipole moments reaches a steady state. Of course, the particles keep moving and interacting after they have reached one $\lambda_{mfp}$, but the anisotropy itself does not change.  From Figure~\ref{fig:aps30PeV} it is possible to see that this steady distribution is not isotropic, yet it possesses a definite structure that is dependent on the nature of the turbulent magnetic field. Consequently, the observed anisotropy is for the most part created in the last $\lambda_{mfp}$ of the particle's trajectory, and it becomes a way to indirectly probe the local ISM.

In Figure $\ref{fig:aps30PeV}$, we can see from the comparison with the observations that the experimental data behaves according to what we would have expected from a lower energy. In the case of 20 TeV, the distribution of power among the higher multipole moments is lower than at the higher energy, i.e., 30 PeV. One of the causes is that the 30 PeV particles interact with perturbations that carry more power than the ones at a lower energy. Therefore, when an interaction process occurs with these perturbations, the higher energy particles are affected more strongly and more small-scale structure is created.

This effect is confirmed in Figure $\ref{fig:apscomparison}$, where the spectrum at the $\lambda_{mfp}$ of energies 750 TeV and 30 PeV is compared with the observations at the median energy of 20 TeV. The distribution at 750 TeV is similar to the observations at 20 TeV; however, for 30 PeV, the distribution of high multipole moments is much higher and flatter than that at lower energy. Therefore, greater small-scale anisotropy is observed at higher energies. This is also easily seen in the maps in Figure~\ref{fig:mapcomparison}.

The flatter angular power spectrum obtained with 30 PeV compared to 750 TeV protons tells us about differences in pitch angle scattering as a function of energy. However it is important to note that the 750 TeV data set corresponds to a gyroradius scale close to the dissipation region of the MHD turbulence inertial range. It is likely that scattering is underestimated, thus preventing a full development of small-scale angular structures in the particle anisotropy. From this point of view, it is reasonable to assume that the resemblance of the 750 TeV proton power spectrum to experimental data should be considered coincidental but still in agreement with the fact that lower energies should have less structure, as mentioned above. On the other hand, the difference of the 30 PeV proton power spectrum with experimental data does not necessarily mean that the fundamental processes responsible for the small-scale anisotropy are overestimated in the present study. The results show an energy dependence in the shape of the angular power spectrum that needs further study, requiring MHD turbulence simulation with a significantly wider inertial range. 

Figure $\ref{fig:apscomparison}$ includes the angular power spectrum resulting from hierarchical decay of angular scales if Liouville's theorem is satisfied, as calculated by~\cite{ahlers_2014}. Here the different curves have been normalized to the dipole of the observational data, as discussed in Section~\ref{sec:aps}, so that only the small-scale structure is relevant. In~\cite{ahlers_2014}, an existing global dipole anisotropy evolves to create higher multipole moments. The difference between that result and the one shown in the present work resides in the fact that here the shape of the distribution is determined by the specific turbulence characteristics of the magnetic field and the gyroradius of the particles. The 750 TeV case appears similar to ~\cite{ahlers_2014}, but since this set is at the damping scale and scattering is likely to be underestimated, we were already expecting less structure to be present.

\cite{ahlers_mertsch_2015} studied the effect of relative diffusion, i.e., from correlated nearby trajectories, as the contribution to the development of small scale anisotropy structures. In this complementary approach, the angular power spectrum is calculated based on the effect that a particle density gradient has on the trajectory topology shaped by homogeneous isotropic magnetic turbulence. In the present work, the relationship between the shape of angular power spectrum and the scattering processes induced by Alfv\'enic, slow, and fast modes in a compressible MHD turbulence were studied. This will make it possible to identify physical properties that shape the angular power spectrum, so that the problem can be inverted, i.e., the angular power spectrum and different cosmic ray energies and masses can be used to probe the properties of the interstellar magnetic turbulence.


\section{Conclusions}
\label{sec:concl}








This work explores the possibility that the local interstellar magnetic field could shape the arrival direction distribution of high-energy cosmic rays. An MHD turbulent magnetic field was used as a propagating medium that resembles the ISM, and particle trajectories with various energies were integrated in this field. To obtain the anisotropy in arrival direction distribution at Earth, the theoretical framework for application of Liouville's theorem was provided, and the theorem was shown to be applicable in this specific case. Nonetheless, there could be cases where the magnetic field varies abruptly, which would ultimately violate the application of the theorem.

The results presented in this work show that small-scale anisotropy arises from the interaction of cosmic rays with the local turbulent magnetic field, as confirmed in the sky maps and angular power spectra.
The angular power spectrum becomes flatter the higher the energy; therefore, experimental data at the 10 TeV scale is expected to be steeper than the numerical calculations presented in this paper. Cosmic rays with less rigidity are more sensitive to the lower power small-scale magnetic perturbations. Since cosmic rays are not composed only of protons, the contribution of heavy ions would yield a steeper angular power spectrum.

The inertial range of our turbulent magnetic field provides a limitation on the lowest energy that could be studied, in this case 750 TeV. Therefore, in the future, it will be necessary to extend this inertial range and sample even lower energies, so that direct comparisons with the observations at 20 TeV will be possible. Still, it is expected that at these TeV energies the effects of the heliosphere should be present as well, which would provide a primary topic for future study and publication. 
Another factor to improve is the number of particles that are propagated. If we were to have at least $10^{7}$ events, it would be possible to obtain a clearer view of the distribution of cosmic rays at Earth.
Investigating how the properties of our local turbulent magnetic field might influence TeV-PeV cosmic ray arrival direction distribution will provide the basis for further exploring the observed anisotropy, and it will open the doors for a better understanding of our local interstellar medium.








\acknowledgments

Many thanks to Markus Ahlers, Juan Carlos D\'iaz V\'elez, and other colleagues at WIPAC, and to Ellen Zweibel, Federico Fraschetti, Martin Pohl, Rahul Kumar and Jungyeon Cho for useful and constructive discussions on the topic. AL acknowledges the support of NSF grant AST 0808118, NASA grant X5166204101, and the NSF-sponsored Center for Magnetic Self-Organization. PD acknowledges support from WIPAC and the U.S. National Science Foundation-Office of Polar Programs. VLB thanks the support of the Brazil-U.S. Physics Ph.D. Student and Post-doc Visitation Program.
This work was partially supported by the Research Experiences for Undergraduates (REU) Program of the National Science Foundation under Award Number AST-1004881.

\newpage
\begin{appendix}

\section{Numerical Approach and Accuracy}
\label{app:a}

Particle trajectories are calculated by integrating the 6D set of equations of motion, eqs.~1 and~\ref{eq:motion2}. As stated in section~\ref{ssec:crprop}, the integration is performed numerically using the Bulirsch-Stoer method, which is considered one of the best known integration algorithms satisfying both high accuracy and efficiency~\citep{press_1986} and widely applied in the literature (e.g.,~\cite{giacalone_jokipii_1999} and~\cite{xu_yan_2013}). The Bulirsch-Stoer integration algorithm, is a known method for numerical calculation of ordinary differential equation solutions, that combines the so-called Richardson extrapolation (to improve the rate of convergence of a sequence) and the modified midpoint method (which advances a vector of dependent variables y(x) from a point x to a point x + H by a sequence of n substeps each of size h). The result is that Bulirsch-Stoer algorithm provides high accuracy with relatively low computational effort. The accuracy of the algorithm is further controlled, during the numerical calculation, by monitoring the local truncation error estimated at each time step. If the relative error is larger than the relative tolerance level of 10$^{-6}$, the step size is adaptively reduced in order to limit the error accumulation in both momentum and spatial coordinates, across the maximum integration time used in this work (corresponding to no more than 10,000 gyrations). Such error accumulation needs to be monitored for each specific problem in which the integration algorithm is used. In particular, the accuracy on the spatial and momentum coordinates were studied.

\begin{figure*}[h!]
\vspace*{-2cm}
\begin{center}
\includegraphics[width=0.45\columnwidth]{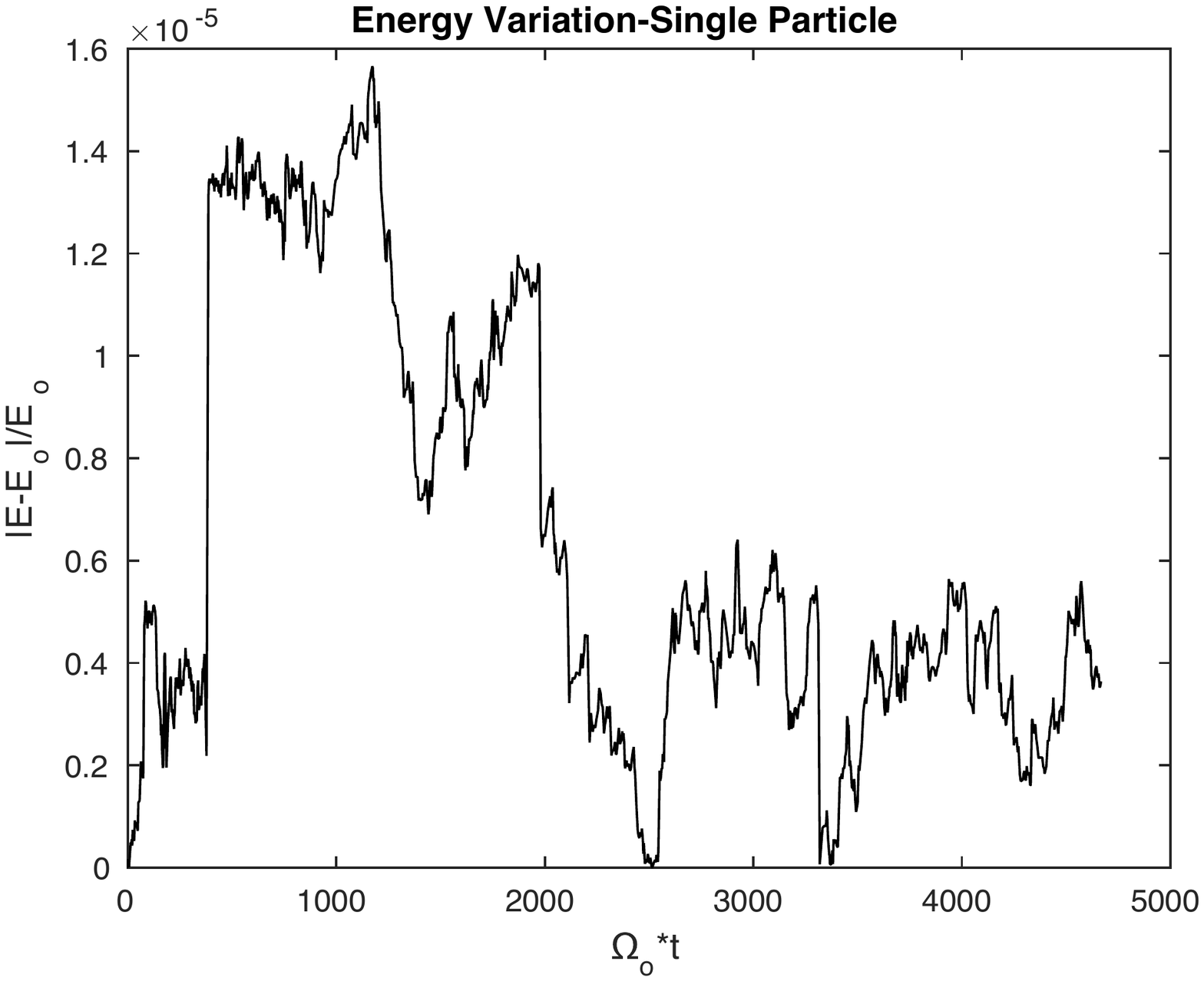}
\includegraphics[width=0.45\columnwidth]{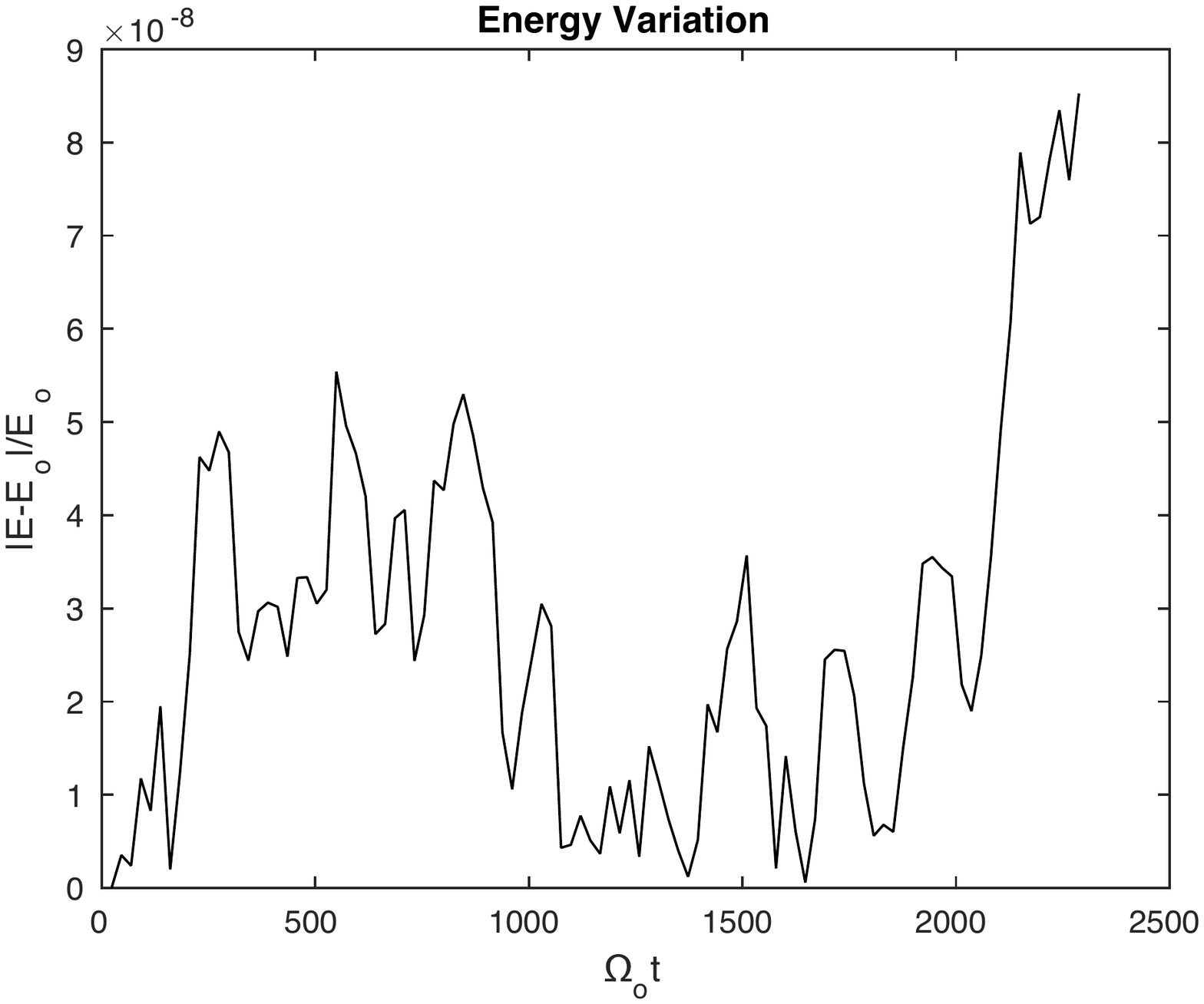}
\end{center}
\vspace*{-2cm}
\caption[Energy Concervation]
{Accuracy of the conservation of energy for a single particle (on the left) and for the average particle sample (on the right) of the 30 PeV set of Table~\ref{tab:sets}.}
\label{fig:appa}
\end{figure*}

The performance of the Bulirsch-Stoer algorithm on the accuracy of momentum coordinates in our numerical calculation can be seen in Figure~\ref{fig:appa}, where the particle energy variation (due to loss of accuracy) from that at $t=0$ is plotted as a function of the number of gyro-orbits $\Omega_0\,t$ (for a single particle and for the average of all particles used in our 30 PeV sample).

The maximum relative mean deviation from perfect energy conservation is found to be about 8.5$\times$10$^{-8}$ for the sample used in this work (100,000 particles in the 30 PeV energy set), although a single particle can reach a violation at the 1.6$\times$10$^{-5}$ level. This precision level in the energy conservation guarantees that particle trajectories are not significantly affected by numerical accuracy limitations, which are found to be marginal under the conditions of this study.

Since the adaptive time step algorithm constrains both spatial and momentum coordinates to the same relative error level, the accuracy in spatial coordinates is $\ll r_L$, even after 10,000 gyrations. Numerical diffusion, therefore, is limited to a level much smaller than Bohm diffusion, which is several orders of magnitude below diffusion induced by the stochastic wandering of magnetic field lines at all scales~\citep{ly_2014}.

The effect on the particle set size was assessed in~\cite{xu_yan_2013}, which used the same integration stepping algorithms as in this work. In that paper, the perpendicular diffusion coefficient becomes stable when the sample size reaches about 1000 particles. The sets used in this study contain 100,000 or more particles (see Table~\ref{tab:sets}), thus minimizing statistical accuracy effects on the global behavior of the ensemble of particles.

\end{appendix}


\end{document}